\documentclass[aps,reprint,citeautoscript,groupedaddress,longbibliography]{revtex4-1}
\pdfoutput=1

\usepackage{amssymb,amsmath,mathtools,graphicx}
\usepackage{xspace}  
\usepackage{siunitx} 

\newcommand{\vek}[1]{\boldsymbol{#1}}          
\newcommand{\dif}{\mathrm{d}}                  

\newcommand{\mean}[1]{\left<#1\right>}
\newcommand{\abs}[1]{\left|#1\right|}
\def\Real{\hbox{I\kern-.1667em\hbox{R}}}

\newcommand{\Ld}{L$_{\text{d}}$\xspace}   
\newcommand{\Lo}{L$_{\text{o}}$\xspace}   
\newcommand{\lmax}{l_{\text{max}}}            
\newcommand{\lstar}{l^*}            
\newcommand{\kstar}{k^*}            
\newcommand{\kmin}{k_{\text{min}}}            

\DeclareMathOperator*{\argmin}{arg\,min}   

\begin{document}
\title{
Phase Diagrams of Multicomponent Lipid Vesicles: \\ Effects of Spherical Topology and Finite Size
}
\author{Yongtian Luo}
\author{Lutz Maibaum}
\email{maibaum@uw.edu}
\affiliation{Department of Chemistry, University of Washington, Seattle, WA 98195 }
\begin{abstract}
We study the phase behavior of multicomponent lipid bilayer vesicles that can exhibit intriguing morphological patterns and lateral phase separation. We use a modified Landau-Ginzburg model capable of describing spatially uniform phases, microemulsions, and modulated phases on a spherical surface. We calculate its phase diagram for multiple vesicle sizes using analytical and numerical techniques as well as Monte Carlo simulations. Consistent with previous studies on planar systems, we find that thermal fluctuations move phase boundaries, stabilizing phases of higher disorder. We also show that the phase diagram is sensitive to the size of the system at small vesicle radii. Such finite size effects are likely relevant in experiments on small, unilamellar vesicles and should be considered in their comparison to theoretical and simulation results.
\end{abstract}
\maketitle

\section{\label{sec:intro}Introduction}

Bilayers made of multiple species of lipid molecules form the underlying structure of biological membranes. An urgent question in membrane biophysics is whether such bilayers are laterally uniform or whether they exhibit spatial inhomogeneities, i.e., regions that differ in local composition either permanently or transiently. The relevance of this question stems from its implications for the spatial distribution of membrane proteins, which is influenced by the local lipid environment and which directly affects fundamental biological processes such as cell signaling.

The ability of multicomponent bilayers to display inhomogeneous lateral structure is vividly demonstrated in fluorescence microscopy experiments on giant unilamellar vesicles (GUVs) that contain three types of lipids: a lipid with high melting temperature, typically a saturated phospholipid such as dipalmitoylphosphatidylcholine (DPPC) or distearoylphosphatidylcholine (DSPC), a lipid with low melting temperature, for example an unsaturated phospholipid like dioleoylphosphatidylcholine (DOPC), and cholesterol. Over a wide range of compositions these ternary systems exhibit a transition from a high-temperature, homogeneous state to a low-temperature, phase-separated state in which the vesicle partitions into two distinct regions: the liquid-ordered (\Lo) phase, rich in saturated lipids and cholesterol, and the liquid-disordered (\Ld) phase that contains mostly unsaturated lipids~\cite{Veatch03,Scherfeld03,Veatch05}. This ability to support multiple distinct fluid phases has been found in many ternary lipid mixtures~\cite{Marsh09} using several experimental techniques including infrared spectroscopy~\cite{Silvius96,Wang01}, NMR~\cite{Veatch04}, FRET~\cite{Zhao07, Heberle10}, small-angle neutron or X-ray scattering (SANS or SAXS)~\cite{Pencer05,Pabst10,Kollmitzer15} in addition to optical microscopy. It has also been found in computer simulations~\cite{Risselada08,Rosetti11,Sodt14,Baoukina17,Pantelopulos17} and theoretical models~\cite{Radhakrishnan05,Almeida09,Putzel11b,Svetlovics12} of mixed bilayers.

The discovery of thermodynamically stable liquid-ordered phases in ternary lipid systems has spurred renewed interest in the lipid raft model of membrane organization~\cite{Simons97,Brown98,Lingwood10}, which predicts the existence of small, transient domains of increased lipid order in biological membranes. While the quest to identify such domains in these much more complex bilayer systems is ongoing, there is clear evidence for \Lo/\Ld phase separation in plasma membrane-derived vesicles~\cite{Baumgart07,Veatch08,Levental15}.

Increasing the chemical complexity of a lipid bilayer can induce additional types of lateral organization not seen in ternary mixtures. For example, upon adding a fourth lipid type with one saturated and one unsaturated tail (sometimes referred to as a hybrid lipid) one finds bilayers containing nanoscopic domains as well as modulated (stripe) phases~\cite{Konyakhina11, Goh13, Heberle13b, Konyakhina13}. While the latter cannot be observed in coarse-grained molecular dynamics simulations~\cite{Ackerman15, He18} due to their limited system size, modulated phases have emerged in Monte Carlo simulations of a discretized surface model of phase-separated membranes that captures differences in the bending moduli of liquid-ordered and liquid-disordered regions~\cite{Amazon13}.

Modulated phases also emerge when a system favors the creation of interfaces between liquid-ordered and liquid-disordered regions in the bilayer. This can be caused, for example, by a coupling between composition and shape fluctuations in asymmetric bilayers~\cite{Shlomovitz13}. Hybrid lipids that accumulate at interfaces between \Lo and \Ld phases and that thereby effectively reduce the interface tension might have a similar effect~\cite{Palmieri13}.

The Landau-Ginzburg (LG) model is an elegant description of a system that exists in a homogeneous, disordered state at high temperatures and that supports multiple coexisting phases at low temperatures~\cite{Goldenfeld,Safran03}. To include the possibility of modulated phases, the LG model was extended by Shlomovitz and Schick to allow for an effectively negative interface tension between liquid-ordered and liquid-disordered domains~\cite{Shlomovitz13}. A mean field analysis of this model revealed a new microemulsion phase in addition to the anticiapted modulated, homogeneous, and coexisting phases. Monte Carlo simulations of this model later revealed that 
the phase diagram changes significantly if the effects of thermal fluctuations are considered~\cite{Shlomovitz14}.

Both the mean field and the finite temperature analysis of the extended LG model in references \onlinecite{Shlomovitz13} and \onlinecite{Shlomovitz14} were performed under the assumption of a planar bilayer patch. Many experiments, however, are performed on vesicle systems that have the topology and usually also the shape of a sphere, and one might wonder to what extent this change in system geometry affects the phase behavior of the model. This question is particularly pertinent for small vesicles such as those used in neutron scattering studies, which are only \SI{60}{nm} in diameter~\cite{Heberle13b}. We have recently shown that the principal observable in such scattering experiments, the static structure factor, can differ significantly between planar and spherical systems~\cite{Luo18a}. Furthermore it is known that even seemingly simple models can generate surprisingly complex spatial patterns on spherical surfaces~\cite{Lavrentovich16}.

For these reasons we study in this work the extended Landau-Ginzburg model on spherical surfaces. The previous results for the planar system will serve as our reference point that we expect to recover in the limit of large sphere radii~\cite{Shlomovitz13, Shlomovitz14}. The model is described in detail in Section~\ref{sec:model}. We analyze the model in three different ways, as outlined in Section~\ref{sec:method}: we study an approximate form of the ground state that can be obtained analytically, the actual ground state obtained from numerical energy minimization, and the finite-temperature behavior of the model using Monte Carlo simulations. Results of these calculations are summarized in Section~\ref{sec:result}. We discuss several general aspects of finite size and fluctuation effects in membrane systems in Section~\ref{sec:discuss}.

\section{\label{sec:model}The Model}

To describe the lateral structure of a multicomponent lipid bilayer we introduce a scalar order parameter field $\phi(\vek{r})$ that allows us to distinguish between different types of local membrane structure. This field is defined on a two-dimensional surface $S$ representing the shape of the bilayer. Our model assigns to each realization of this field a (Landau) energy through the functional~\cite{Shlomovitz14}
\begin{align}
\begin{split}
\label{eqn0}
E[\phi(\vek{r})]=\int\dif S\,\Big\{
&
\frac{\alpha}{2}\abs{\phi(\vek{r})}^2
+\frac{b}{4}\abs{\phi(\vek{r})}^4   \\
& +\frac{\gamma}{2}\abs{\nabla \phi(\vek{r})}^2
+\frac{\epsilon}{2}\abs{\nabla^2 \phi(\vek{r})}^2  
\Big\} .
\end{split}
\end{align}
The first three terms in this expression are the famous Landau-Ginzburg model for continuous order--disorder transitions~\cite{Goldenfeld,Safran03}. Its principal features can be obtained by minimizing the energy with respect to the field $\phi(\vek{r})$. If all three parameters $\alpha$, $b$ and $\gamma$ are positive, then the lowest energy configuration is the uniform field $\phi(\vek{r}) = 0$. We interpret this state as the homogenous, well-mixed phase of the system. Making $\alpha$ negative we obtain two degenerate, uniform ground states, $\phi(\vek{r}) = \pm \sqrt{-\alpha/b}$. These states with non-zero order parameter values are at coexistence with each other, and we interpret them as the liquid-disordered and liquid-ordered phase, respectively. In absence of constraints the system will adopt one or the other at random, but if, for example, a conservation law determines the area fractions of these two phases the system will partition into distinct \Ld and \Lo domains, separated by an interface of characteristic width $\sqrt{-2 \gamma/\alpha}$~\cite{Safran03}. The line tension, i.e., the energy per unit length of this interface, is
\begin{equation}
\sigma=\sqrt{-\frac{8\gamma\alpha^3}{9b^2}} .
\end{equation}

Extending this model to also describe modulated phases that are rich in domain boundaries, Shlomovitz and coworkers considered the case of negative $\gamma$, motivated by previous work on asymmetric lipid bilayers~\cite{Schick12, Shlomovitz13}. This requires the introduction of the final term in \eqref{eqn0} with a positive parameter $\epsilon$ to maintain thermodynamic stability. In addition to the single homogeneous phase and the region of phase coexistence, this extended model can also sustain both a modulated and a microemulsion phase~\cite{Shlomovitz14}.

While this analysis was performed on planar bilayer systems, we now consider spherical vesicles of radius $R$. We rewrite~\eqref{eqn0} in spherical coordinates,
\begin{align}
\begin{split}
\label{eqn1}
E[\phi(\theta,\varphi)]=  R^2 \int & \dif \Omega  \,\Big\{
\frac{\alpha}{2}\abs{\phi(\theta,\varphi)}^2
+\frac{b}{4}\abs{\phi(\theta,\varphi)}^4   \\
& +\frac{\gamma}{2}\abs{\nabla \phi(\theta,\varphi)}^2
+\frac{\epsilon}{2}\abs{\nabla^2 \phi(\theta,\varphi)}^2  
\Big\}  ,
\end{split}
\end{align}
where the integral is over all solid angles $\Omega$, the field $\phi$ is a function of the inclination angle $\theta$ and the azimuthal angle $\varphi$, and the derivatives are to be taken within the spherical surface. For example, the Laplace operator becomes
\begin{equation}
 \nabla^2=\frac{1}{R^2\sin^2\theta}\frac{\partial^2}{\partial\varphi^2}+\frac{1}{R^2\sin\theta}\frac{\partial}{\partial\theta}\left(\sin\theta\frac{\partial}{\partial\theta}\right) \label{eq:laplace}.
\end{equation}

For numerical calculations it is convenient to expand the field $\phi$ in a basis of spherical harmonic (SH) functions,
\begin{equation}
\label{eqn2}
 \phi(\theta,\varphi)=\sum_{l=0}^{\infty}\sum_{m=-l}^{l}w_{l,m}Y_{l,m}(\theta,\varphi) .
\end{equation}
The complex coefficients $w_{l,m}$ can be obtained from the SH transform of $\phi$,
\begin{equation}
\label{eqn3}
 w_{l,m}=\int\dif\Omega\,Y_{l,m}^*(\theta,\varphi)\phi(\theta,\varphi) .
\end{equation}
Because the field is real-valued the SH coefficients satisfy the relationship
\begin{equation}
\label{eqn_ex}
 w_{l,m}^*=(-1)^m w_{l,-m} .
\end{equation}

Expressed in terms of these coefficients the energy \eqref{eqn1} becomes (see Appendix~\ref{append1} for details)
\begin{widetext}
\begin{equation}
\label{eqn4}
\begin{split}
E
&=\frac{R^2}{2}\sum_{l,m}\left\{\alpha+\frac{\gamma}{R^2}l(l+1)+\frac{\epsilon}{R^4}[l(l+1)]^2\right\}\lvert w_{l,m}\rvert^2\\
&\quad+\frac{bR^2}{4}\sum_{l,m}\left\lvert\sum_{l_1,m_1}\sum_{l_2,m_2}\sqrt{\frac{(2l_1+1)(2l_2+1)(2l+1)}{4\pi}}\begin{pmatrix}l_1&l_2&l\\0&0&0\end{pmatrix}\begin{pmatrix}l_1&l_2&l\\m_1&m_2&m\end{pmatrix}w_{l_1,m_1}w_{l_2,m_2}\right\rvert^2 .
\end{split}
\end{equation}
\end{widetext}
Here we introduced the shorthand notation $\sum_{l,m}$ for sums over SH indices such as in~\eqref{eqn2}, and we see the emergence of the Wigner 3j symbols in the expression for the term that is quartic in the order parameter field.

\section{\label{sec:method}Computation and Simulation Methods}

We are interested in the phase behavior of the model~\eqref{eqn4} as a function of the material parameters $\alpha$, $\gamma$, $\epsilon$, and $b$ as well as the system's radius $R$.  We work in dimensionless units, and limit ourselves to $\epsilon = b = 1$ without loss of generality~\cite{Shlomovitz14}. We obtain phase diagrams by either energy minimization or Monte Carlo simulations for parameter values $-3 \leq \alpha \leq 3$ and $-3 \leq \gamma \leq 1$, which we explore with a resolution of $\Delta\alpha = \Delta\gamma = 0.1$. For each set of parameters we consider three system sizes, $R = 1$, 3, and 10, the last one being large enough to approach the behavior of a planar system.

We perform three distinct calculations, as outlined below. The first is an energy minimization using a very limited basis set of only a single spherical harmonic function. The second is an energy minimization in a much larger basis. The third is a Monte Carlo simulation that allows to sample the thermal equilibrium ensemble of order parameter fields.

\subsection{\label{subsec:singlemode}Single-Mode Energy Minimization}

In this approximation we include only a single term in the SH expansion~\eqref{eqn2},
\begin{equation}
\phi(\theta)=w_{l,0}Y_{l,0}(\theta) .
\end{equation}
We limit ourselves to a term with $m=0$, which implies that the field $\phi$ is independent of the azimuth $\varphi$. In this case the expression~\eqref{eqn4} for the energy can be written as
\begin{equation}
 \label{eqn5}
 \begin{split}
E=
&\frac{R^2}{2}\left\{\alpha+\frac{\gamma}{R^2}l(l+1)+\frac{\epsilon}{R^4}[l(l+1)]^2\right\}w_{l,0}^2\\
&+\frac{bR^2}{4} \left[ \sum_{l'=0}^{2l}\frac{(2l+1)^2(2l'+1)}{4\pi}\left\lvert\begin{pmatrix}l&l&l'\\0&0&0\end{pmatrix}\right\lvert^4  \right] w_{l,0}^4 ,
\end{split}
\end{equation}
a quartic polynomial in $w_{l,0}$. We minimize the energy with respect to the spherical harmonic degree $l$ and to the value of the coefficient $w_{l,0}$:
\begin{equation}
( \lstar, w_{\lstar\!,0}) = \argmin_{l , w_{l,0} } E .
\end{equation}

There are three possible outcomes:
\begin{enumerate}
\item If the energy is minimal at $w_{\lstar\!,0} = 0$, then the ground state of the system is the uniform field $\phi = 0$, which has energy $E=0$. This state represents the homogeneous fluid.
\item If the energy is minimal at $w_{\lstar\!,0} \neq 0$ and $\lstar=0$,  then there are two degenerate ground states with energy $E=-\pi\alpha^2 R^2/b$. Both are uniform fields with $\phi = \pm \sqrt{-\alpha/b}$. These two phases are at coexistence with each other.
\item If the energy is minimal at $w_{\lstar\!,0} \neq 0$ and $\lstar>0$, then the ground state consists of $l+1$ stipe domains of alternating positive and negative values of the order parameter, representative of the modulated phase.
\end{enumerate}
For every parameter pair $(\alpha,\gamma)$ we compare the energies for each scenario, considering values of $l$ up to 21. Due to the simple form of $\eqref{eqn5}$ both the assignment of the minimum energy phase and the identification of phase boundaries can be done analytically.

\subsection{\label{subsec:fullmode}Full-Mode Energy Minimization}

Here we include all spherical harmonic functions with an index up to $\lmax=21$ and order $m=0$ in the expansion~\eqref{eqn2}:
\begin{equation}
\phi(\theta)=\sum_{l=0}^{l_{\text{max}}}w_{l,0}Y_{l,0}(\theta)
\end{equation}
In this case the energy~\eqref{eqn4} becomes
\begin{widetext}
\begin{equation}
\label{eqn6}
\begin{split}
E=
&\frac{R^2}{2}\sum_{l=0}^{l_{\text{max}}}\left\{\alpha+\frac{\gamma}{R^2}l(l+1)+\frac{\epsilon}{R^4}[l(l+1)]^2\right\}w_{l,0}^2\\
&+\frac{bR^2}{4}\sum_{l=0}^{2l_{\text{max}}}\left[\sum_{l_1=0}^{l_{\text{max}}}\sum_{l_2=0}^{l_{\text{max}}}\sqrt{\frac{(2l_1+1)(2l_2+1)(2l+1)}{4\pi}}\left\lvert\begin{pmatrix}l_1&l_2&l\\0&0&0\end{pmatrix}\right\lvert^2 w_{l_1,0}w_{l_2,0}\right]^2  .
\end{split}
\end{equation}
\end{widetext}

At every point $(\alpha, \gamma)$ in the phase diagram we minimize this energy numerically with respect to the SH coefficients $\{w_{l,0}\}$. The last contribution, which originates from the quartic term in~\eqref{eqn4}, is evaluated using the SHTOOLS spherical harmonics library~\cite{SHTOOLS34}. We use both the  Broyden-Fletcher-Goldfarb-Shanno (BFGS) and the Sequential Least Squares Programming (SLSQP) optimization algorithm as implemented in the SciPy scientific programming library, and choose the configuration with the lowest energy. To further refine the obtained ground states we perform multiple sweeps through the parameter space, seeding the energy minimization with ground states obtained for nearby parameter values.

The ground state configurations are classified into the same three categories as before. If all $w_{l,0}$ are zero, then the system is in the homogeneous phase. Otherwise we identify the index $\lstar$ for which  $\abs{w_{l,0}}$ is largest. If that dominant index is $\lstar=0$ then the system is at coexistence between two uniform phases. If, one the other hand, $\lstar \geq 1$ then the system is in the modulated phase.

\subsection{\label{subsec:MonteCarlo}Monte Carlo Simulations}

We perform Markov Chain Monte Carlo simulations to sample the equilibrium distribution of the order parameter field. Here we include all terms of the expansion~\eqref{eqn2}, including those with non-zero $m$, up to $\lmax = 21$. 

The independent degrees of freedom are the real parts $c_{l,m}$ and imaginary parts $s_{l,m}$ of the SH coefficients $w_{l,m}$ with $m \geq 0$, with the exception of $s_{l,0}$ which is always zero for real order parameter fields. The coefficients for negative $m$ are uniquely determined by the symmetry relation~\eqref{eqn_ex}.

As before we use the SHTOOLS library to evaluate the quartic term in the energy~\cite{SHTOOLS34}. In each Monte Carlo step we choose one of these degrees at random, and propose a change drawn from a normal distribution with standard deviation $0.1\times\sqrt{T/(\alpha R^2+\gamma+\epsilon/R^2)}$ if $\alpha R^2+\gamma+\epsilon/R^2>0$, or $0.1$ otherwise. Here $T = 0.1$ is the temperature of the system. This trial move is then accepted or rejected according to the Metropolis criterion. A typical simulation required $10^6\times(\lmax+1)^2$ steps to converge.

From these calculations we obtain the equilibrium averages $\mean{c_{l,m}}$ and variances $\mean{(\delta c_{l,m})^2} = \langle c_{l,m}^2\rangle - \mean{c_{l,m}}^2 $, and similar for $s_{l,m}$. For each value of the index $l$ there exist $2l+1$ independent real-valued coefficients that have to be considered for the assignment of thermodynamic phases. If the average value of all coefficients is zero then the system is in the homogeneous phase. If the average value of all coefficients with $l \geq 1$ is zero and $\mean{c_{0,0}} \neq 0$ then the system is at coexistence. If there is a peak in average coefficient values at a non-zero SH index $\lstar$ then the system is in the modulated phase.

Because Monte Carlo simulations capture the effects of thermal fluctuation we can use them to identify another phase, the microemulsion. It is characterized by significant fluctuations at a specific length scale. In planar systems it can be detected by a peak at non-zero wave vectors in the static structure factor~\cite{Schick12, Shlomovitz13, Shlomovitz14}. For the spherical systems considered here we use the analogous condition for the SH coefficients: we identify a microemulsion by a peak in the variances of either $c_{l,m}$ or $s_{l,m}$ at a non-zero index $\lstar$, while the means of the coefficients remain zero~\cite{Luo18a}.

\begin{figure*}
\centering
\includegraphics[width=\textwidth]{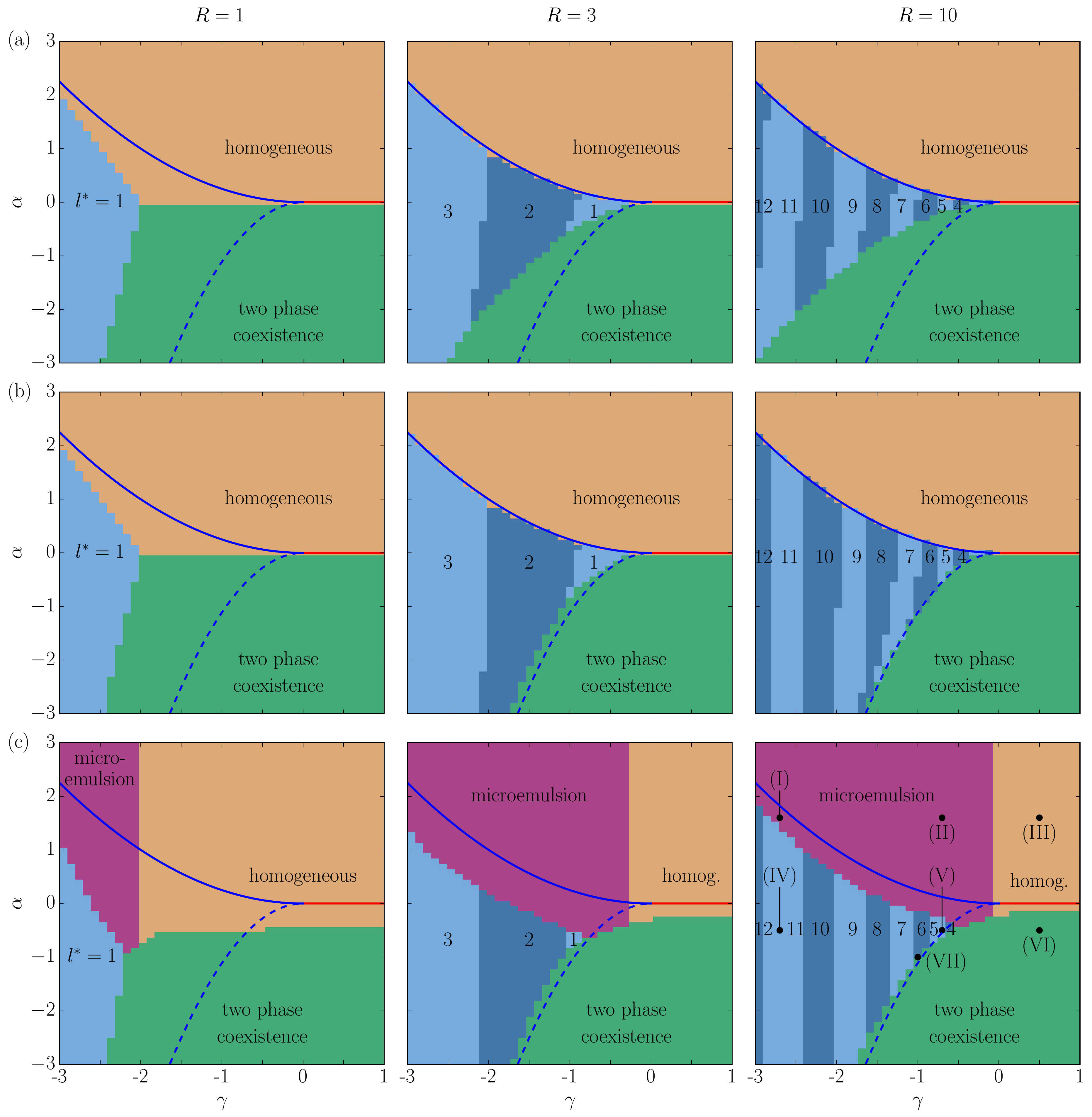}
\caption{
Phase diagrams of the Landau-Ginzburg model~\eqref{eqn0} on the spherical surfaces of radii $R=1$, 3, and 10, as obtained using (a) the single mode ground state calculation, (b) the full mode ground state calculation, and (c) Monte Carlo computer simulations. At fixed $\epsilon=b=1$ the homogenous phase occupies the first quadrant and the coexistence region the fourth quadrant in the $(\alpha,\gamma)$ plane. The modulated phase, which lies at sufficiently negative $\gamma$ is further subdivided into regions of distinct values of $\lstar$. Inclusion of thermal fluctuations gives rise to the microemulsion phase. Also included in each diagram are the phase boundaries obtained for the corresponding planar system (Appendix~\ref{append2}). The seven state points indicated in the bottom right panel are discussed in Figures~\ref{fig2} and \ref{fig3}.
}
\label{fig1}
\end{figure*}

\section{\label{sec:result}Results}

We present in Figure~\ref{fig1} the phase diagrams obtained using each of the three methods discussed in the previous section for three system sizes $R=1$, 3, and 10. As a common point of reference we include in each diagram the phase boundaries of the planar system obtained by mean field theory~\cite{Shlomovitz14} for comparison. The derivation of these curves is summarized in Appendix~\ref{append2}.

\subsection{\label{subsec:singlemoderesults}Single-Mode Energy Minimization}

The top panel of Figure~\ref{fig1} shows the phase diagram in the ($\alpha$,$\gamma$) plane obtained by the single mode energy minimization method of Section~\ref{subsec:singlemode}. As expected from the regular Landau-Ginzburg model, at positive $\gamma$ we find a single homogenous fluid at $\alpha > 0$ and two coexisting fluids at $\alpha < 0$. Because these phases are spatially uniform we obtain the exact ground state despite the restriction of a singe spherical harmonic function as the basis set.

These phases extent into the region of negative $\gamma$. However, for sufficiently negative values there is a transition to the modulated phase, characterized by a positive value of the SH index $\lstar$. In this case the order parameter field $\phi$ consists of $\lstar + 1$ parallel stripes of alternating positive and negative values. The width of these stripes is determined by the material parameters, and it is therefore not surprising that on larger spheres we find a greater number of stripes at the same values of $\alpha$ and $\gamma$.

To estimate the number of stripes we use a result from the planar system~\cite{Shlomovitz14}: in the modulated phase the dominant contribution to the order parameter field has a wave vector of $\kstar =\sqrt{-\gamma/(2\epsilon)}$ (see Appendix~\ref{append2} for a brief derivation). The width of a single stripe is therefore approximately $\lambda = \pi / \kstar$. A sphere of radius $R$ can accommodate roughly $\pi R / \lambda$ such stripes. A reasonable estimate of the parameter $\lstar$ in the modulated region is therefore $\lstar \approx \pi R / \lambda - 1 = R \sqrt{- \gamma/(2 \epsilon)} - 1$.

The exact value of the index $\lstar$ and the boundaries between modulated phases containing different numbers of stripes are obtained by minimizing~\eqref{eqn5}. By equating this expression for adjacent values of $l$ one obtains the boundaries analytically, which are linear but not entirely vertical, as shown in Figure~\ref{fig1}(a). Also linear are the boundaries between the homogenous phase, the coexistence region, and the modulated phase for each value of $\lstar$.

Comparing the phase diagram to that of the planar system we see several noteworthy differences. At small system sizes the triple point is significantly shifted towards negative $\gamma$ from its location at $(\alpha=0, \gamma=0)$ in the planar system. The location and shape of the modulated--homogeneous and modulated--coexistence boundaries are also changed. As the radius increases from 1 to 10, the former phase boundary approaches that of the planar system, while the latter does not. This indicates that the order parameter field is well described by a single spherical harmonic mode in the modulated phase close to the transition towards the homogenous fluid, but not near the phase coexistence region. There the single mode approximation overestimates the energy of the modulated phase.

\subsection{Full-Mode Energy Minimization}

Panel (b) of Figure~\ref{fig1} shows the phase diagram obtained by minimizing the energy with respect to a larger basis set that includes spherical harmonic functions $Y_{l,0}$ up to $\lmax=21$. We find that this phase diagram shares many features with that obtained in the single mode approximation. Nevertheless there are also differences. First, the boundaries within the modulated phase that separate regions with different numbers of stripes appear slightly more vertical than in panel (a). More importantly, we find that the boundary between the modulated phase and the coexistence region now approaches that of the planar system at large system sizes. This confirms our expectation that the phase behavior on a sufficiently large sphere is similar to that of the planar system, the latter acting like the $R \rightarrow \infty$ limit of the former.

\subsection{Monte Carlo Simulations}

The introduction of thermal fluctuations causes significant changes to the phase diagram, as shown in Figure~\ref{fig1}(c). The microemulsion emerges within the homogeneous region, characterized by the ground state ($\phi=0$) but a peak in the variance of a SH coefficient with non-zero index $\lstar.$ It is separated from the homogeneous phase by a vertical boundary called the Lifshitz line, which for large $R$ lies at $\gamma=0$ as in the planar system~\cite{Shlomovitz13,Shlomovitz14} but that shifts toward negative $\gamma$ at small system sizes.

For all values of $R$ the triple point at the intersection of modulated, homogeneous, and coexisting phases in panels (a) and (b) splits into a line, opening up space for a direct transition from the microemulsion to the phase coexistence region. The transitions between modulated and microemulsion phases as well as the boundary between the coexistence region and the homogeneous phase are shifted downward toward smaller $\alpha$, which can be explained by thermal fluctuations stabilizing the less ordered phases. These observations are consistent with previous work on planar systems~\cite{Shlomovitz14}.

\begin{figure}
\centering
\includegraphics[width=\columnwidth]{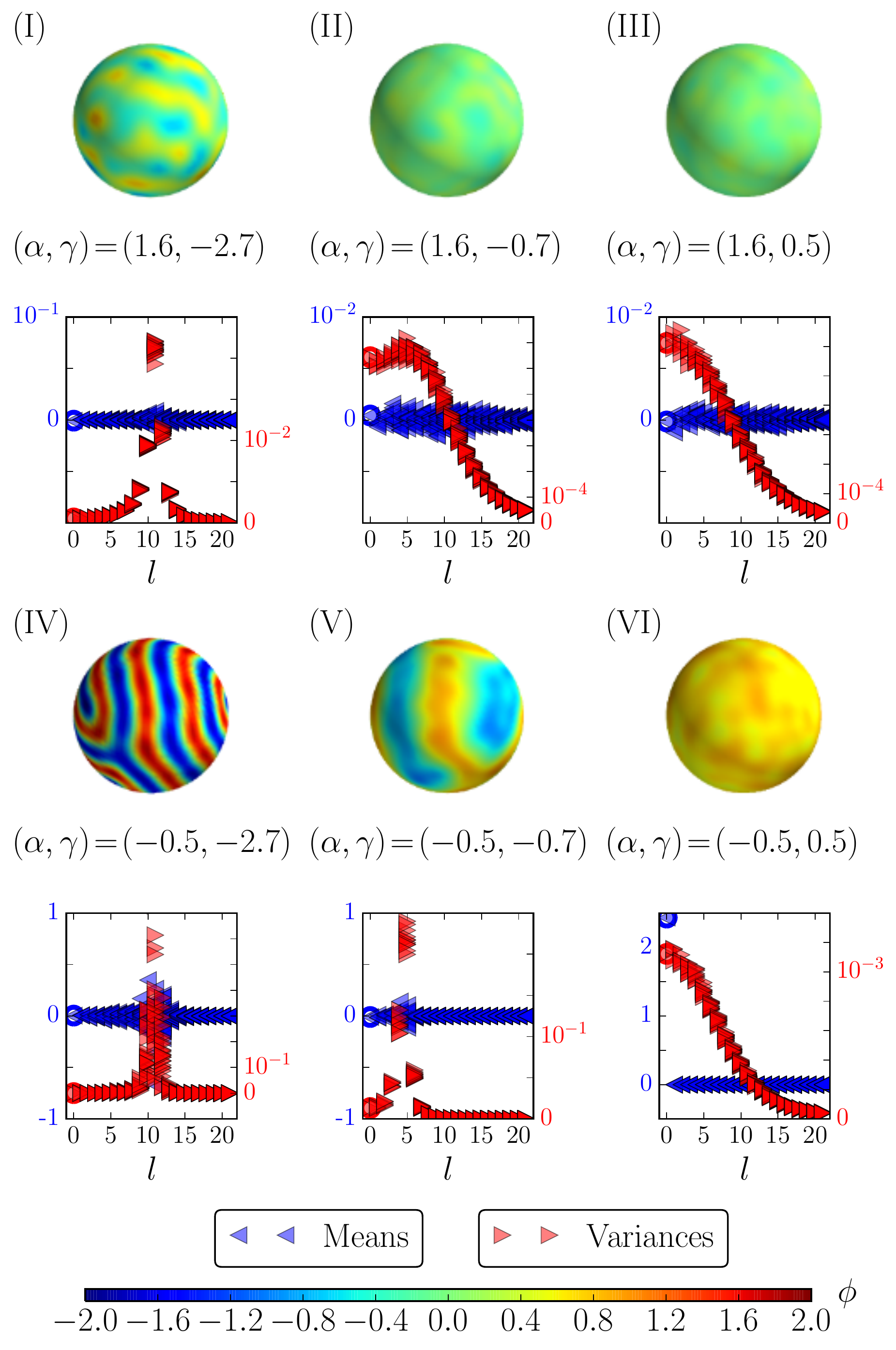}
\caption{
Results from Monte Carlo simulations for $R=10$ at states (I)-(VI) shown in Figure~\ref{fig1}(c). Shown is a snapshot of the order parameter field $\phi$ as well as the means and variances of the spherical harmonic coefficients $c_{l,m}$ and $s_{l,m}$; the values at $l=0$ are highlighted by a circle. These spectra were used to assign to each state the thermodynamic phase: the microemulsion (I and II), the homogeneous fluid (III), the modulated phase (IV and V), and one of the coexisting homogeneous fluids (VI).
}
\label{fig2}
\end{figure}

To illustrate the differences between the various phases and the range of spatial patterns that our model describes we show in Figure~\ref{fig2} snapshots from the Monte Carlo simulations at six different thermodynamic conditions, marked as points (I)--(VI) in Figure~\ref{fig1}(c). These snapshots were obtained by computing the inverse spherical harmonic transformation~\eqref{eqn2} of the coefficients $w_{l,m}$. Also shown in Figure~\ref{fig2} are the means and variances of the degrees of freedom $c_{l,m}$ and $s_{l,m}$ that were used to assign the thermodynamic phases. As discussed in Section~\ref{subsec:MonteCarlo} there are $2l+1$ such degrees for each index $l$.

The points (I) and (II) fall within the microemulsion region of the phase diagram. The simulation snapshots show that the order parameter field $\phi$ is zero on average, but also the presence of thermal fluctuations. The latter can be quantified by considering the spectra of SH coefficients. While the means of $c_{l,m}$ and $s_{l,m}$ are essentially zero for all $l$, there is a peak at $\lstar=11$ (I) or at $\lstar=5$ (II) in their variances that is characteristic for the microemulsion phase. This peak is much sharper in system (I) that is close to the boundary to the modulated phase. Point (III) shows a behavior similar to the first two. However, the variances of the SH coefficients are now monotonically decreasing with $l$, which together with the zero means identifies this point as belonging to the homogeneous phase.

Points (IV)--(VI) show data for the same $\gamma$-values as points (I)-(III), but at much smaller $\alpha=-0.5$. Systems (IV) and (V) show a characteristic stripe pattern, the orientation of which is randomly established over the course of the simulation. The stripe pattern is not perfectly regular but instead contains defects. These stripes are much sharper in (IV) than in (V), the latter being close to the boundaries to both the microemulsion and two-phase coexistence region. The observed surface patterns can be detected in the means and variances of the SH coefficients: both have a peak at non-zero values $\lstar$, which identifies these states as being part of the modulated phase. The values of $\lstar$, 11 for (IV) and 5 for (V), are the same as those found for points (I) and (II), which gives support to the argument put forth in Section~\ref{subsec:singlemoderesults} that $\lstar$ should depend on $\gamma$ but not on $\alpha$. That the length scales of the modulated and the microemulsion phases are independent of $\alpha$ is also seen in planar systems (Appendix~\ref{append3}).

Finally, point (VI) shows the system in one of the coexisting uniform phases with average order parameter $\pm\sqrt{-\alpha/b}$. In this particular simulation the system evolved toward the phase with positive $\phi$. This behavior is reflected in the SH coefficients: the average value of the $l=0$ mode, $\mean{w_{0,0}}$, is close to its theoretical value $\sqrt{- 4\pi \alpha/b} \approx 2.5$, while all other coefficients have an average of zero. The variance, on the other hand, is a monotonically decreasing function of $l$, as expected for a uniform fluid.

\begin{figure}
\centering
\includegraphics[width=\columnwidth]{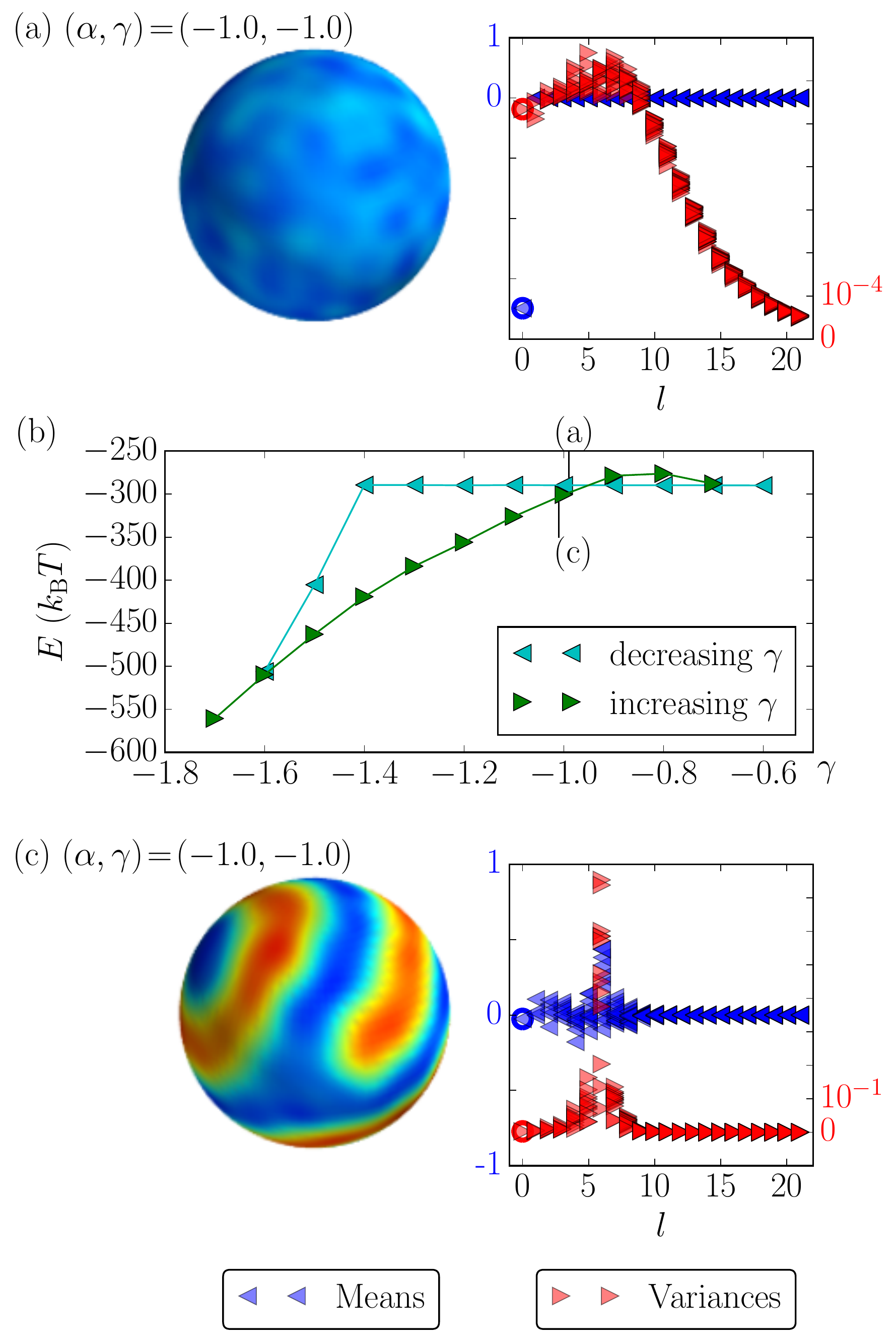}
\caption{
Demonstration of hysteresis and metastability when crossing the boundary between the modulated phase and the coexistence region. Shown are simulation snapshots and the means and variances of spherical harmonic coefficients at $(\alpha,\gamma)=(-1,-1)$ for $R=10$, point (VII) in Figure~\ref{fig1}, obtained from Monte Carlo simulations that were initiated from equilibrium states at greater (a) and smaller (c) values of $\gamma$. The phases of the initial states (a microemulsion at coexistence and the modulated phase, respectively) persist throughout the simulations. Traversing the phase diagram along a horizontal path in both directions shows that the originally stable phases persist beyond the phase boundary (b).
}
\label{fig3}
\end{figure}

To demonstrate some of the difficulties that one encounters when exploring the phase diagram and to study the transition from the modulated phase into the coexistence region in more detail we show in Figure~\ref{fig3} additional data for the point (VII) of Figure~\ref{fig1}(c), which is located at $(\alpha,\gamma)=(-1,-1)$. Shown are the results of two separate simulations of the same thermodynamic state that were started from different initial configurations: the one shown in panel (a) was seeded with a configuration obtained at a larger value of $\gamma$, from inside the coexistence region, while that in panel (c) originated in the modulated phase at smaller $\gamma$.

The data shown in (a) indicates that the system is a microemulsion, as identified by a peak at a non-zero index $\lstar$ in the variances of the spherical harmonic coefficients. However, the average value of the order parameter is not zero but $-\sqrt{- 4\pi \alpha/b} \approx -3.5$. This shows that the Lifshitz line, which  separates the microemulsion from the homogeneous fluid, can be extended to negative $\alpha$, and that the two coexisting phases to the left of the Lifshitz line are in fact microemulsions for $\alpha < 0$.

Unless the point (VII) lies exactly on the boundary, the fact that both the modulated phase and the microemulsion phase are stable even over long simulation time scales shows that one of these two phases is metastable while the other one is stable. To find out which is the thermodynamically stable phase we define a path in the phase diagram that crosses the boundary, and monitor the system as the parameters are changed along this path in both directions.

Figure~\ref{fig3}(b) shows the results of these calculations. We begin at $(\alpha,\gamma)=(-1,-0.6)$, which is well within the coexistence region. Following a path of slowly decreasing $\gamma$ we find that the energy of the system remains nearly constant until we reach $\gamma=-1.5$, where we see a sharp drop in the energy. This drop corresponds to the transition to the modulated phase. Reversing the path by increasing $\gamma$ we find a smooth rise in energy up to $\gamma=-0.8$, where the energy is larger than that of the microemulsion at the same state point. Increasing $\gamma$ even further we eventually find the transition to the stable microemulsion phase. Based on this data we place the boundary of the coexistence region between $\gamma=-1$ and $\gamma=-0.9$, and therefore assign point (VII) to the modulated phase.

The metastability and hysteresis are signatures of a first-order phase transition, as pointed out previously for the planar system~\cite{Shlomovitz14,Sapp14}. This is to be contrasted to the transition from the modulated phase to the microemulsion at positive $\alpha$, which appears to be continuous. In that case there is no change in the average value of the order parameter across the transition, whereas in the case considered in Figure~\ref{fig3} the average changes discontinuously from zero in the modulated phase to $\pm\sqrt{-\alpha/b}$ in one of the coexisting microemulsion phases.

\section{\label{sec:discuss}Discussion}

Our results demonstrate that the principal observations of the role of thermal fluctuations on the phase diagram in planar systems~\cite{Shlomovitz14} also apply in spherical systems. The most striking effect, which is apparent when comparing panels (b) and (c) of Figure~\ref{fig1}, is the splitting of the triple points that leads to a new phase boundary that allows a direct transition between the microemulsion and the phase coexistence region. Another important consequence of thermal fluctuations is the overall shift of the modulated--microemulsion and coexistence--homogeneous boundaries toward lower $\alpha$, which is a ramification of the higher entropy of the more disordered phases.

More importantly, our results show that the finite size can have a significant effect on the phase behavior of multicomponent bilayer systems, as demonstrated in the three phase diagrams shown in Figure~\ref{fig1}(c). It is reassuring that in the limit of large radius $R$ the phase diagram resembles that of the planar system despite the difference in mathematical representation of the model (spherical harmonics vs. plane wave basis set). When decreasing the system size, several changes are discernible. One is the decrease in the value of $\lstar$ in the modulated phase. As discussed in detail in Section~\ref{subsec:singlemoderesults} this observation has an intuitive explanation: the width of the stripes in the modulated phase is determined by material parameters, especially $\gamma$ and $\epsilon$, and the number of such stripes that can be accommodated by the system is proportional to its radius.

Another apparent change is the shift of the Lifshitz line, which separates the microemulsion from the homogenous fluid, toward negative $\gamma$ as the size of the sphere becomes smaller. The magnitude of this shift can be estimated by considering an approximation to the full model~\eqref{eqn4} that omits the quartic part, effectively setting the parameter $b$ to zero. This is permissible in the region of positive $\alpha$ where thermodynamic stability of the model is not affected. Because the remaining terms in the energy are quadratic in the SH coefficients $w_{l,m}$ one can analyze this model analytically, and one finds that their means are zero and their variances~\cite{Luo18a} are
\begin{equation}
\langle\lvert w_{l,m}\rvert^2\rangle=\frac{k_\text{B} T}{\alpha R^2+\gamma l(l+1)+\epsilon[l(l+1)]^2/R^2}. \label{eq:SHgaussian}
\end{equation}
If all parameters are positive then this expression is monotonically decreasing with index $l$, and the system is therefore a homogeneous fluid. Decreasing $\gamma$ below zero, a peak at non-zero index $l$ will eventually form. By equating the variances of the $l=0$ and the $l=1$ terms we find that this first occurs when
\begin{equation}
\label{eqn7}
\gamma=-\frac{2\epsilon}{R^2} .
\end{equation}
This approximate argument yields $\gamma=-2$ for $R=1$, $\gamma=-2/9$ for $R=3$, and $\gamma=-1/50$ for $R=10$ for the location of the Lifshitz line, in agreement with the observed phase behavior.

Another interpretation of~\eqref{eqn7} is that a sphere must have at least a radius of $\sqrt{-2 \epsilon/\gamma}$ in order to realize a microemulsion phase induced by a negative $\gamma$. We have recently shown that the same threshold applies for the detection of a microemulsion in scattering experiments that measure the structure factor of a spherical vesicle embedded in three-dimensional Euclidean space~\cite{Luo18a}. It is heartening that these two approaches yield the same result, and it indicates that the finite size effects discussed here are measurable experimentally.

The dependence of the phase diagram on system size is not limited to spherical vesicles, but in principle can also arise in planar systems such as the one studied previously~\cite{Shlomovitz13,Shlomovitz14}. In that case one uses plane waves instead of spherical harmonic function as a basis set in the expansion of the order parameter field, and typically considers the wave vector as a continuous variable. This, however, introduces the assumption that the system is infinitely large. For a finite system the set of allowed wave vectors is discrete, which causes changes to the phase diagram similar to those reported in this work. We include in Appendix~\ref{append3} a brief overview of these finite size effects in planar systems for the interested reader.

In this work we have limited ourselves to vesicles of fixed spherical geometry. Additional effects arise if the system can deform, which can induce a coupling between bilayer shape and composition. Those are readily observed in experiments~\cite{Baumgart03,Nickels15,Shimobayashi16}, and have been the focus of several  theoretical~\cite{Kawakatsu93,Taniguchi94,Lavrentovich16} and computational~\cite{Taniguchi96,Amazon13,Amazon14,Vidal14,Shimobayashi16} investigations. If the bilayer deformations are small then the effects of this coupling can be absorbed in the parameters of the Landau-Ginzburg model discussed here~\cite{Schick12, Shlomovitz13}. If they are not small, however, an explicit description of the flexible vesicle shape will be necessary.

\section{\label{sec:conclude}Conclusion}

Their finite size is an important characteristic of lipid bilayer vesicles, and one that should be taken into account when exploring their phase diagram. We have shown that both phase boundaries and surface morphologies depend on the radius of a spherical vesicle. While the properties of large vesicles are similar to those of infinitely large, planar membrane patches, small vesicles can behave significantly differently. It is therefore prudent to consider the system size dependence of lipid systems in both theory and experiment.

Our conclusions are based on the analysis of a modified Landau-Ginzburg model that has previously been used to describe planar lipid bilayers. Our results for spherical systems show that the influence of thermal fluctuations on the phase diagram is the same in both geometries: we find the same overall shift of the boundaries between ordered and disordered phases and a splitting of the triple point that creates a boundary between the microemulsion and coexistence regions.

\acknowledgments
This work was facilitated through the use of advanced computational, storage, and networking infrastructure provided by the Hyak supercomputer system at the University of Washington.

\appendix
\section{\label{append1}Energy Expressed in Spherical Harmonic Coefficients}

Here we show how the expansion~\eqref{eqn2} of the order parameter field allows us to write the energy~\eqref{eqn1} in the form~\eqref{eqn4}. Using the orthonormality of the spherical harmonic functions~\cite{DLMF},
\begin{equation}
\int\dif\Omega\,Y_{l,m}^*(\theta,\varphi)Y_{l',m'}(\theta,\varphi)=\delta_{l,l'}\delta_{m,m'} ,
\end{equation}
together with the fact that the $Y_{l,m}(\theta,\varphi)$ are eigenfunctions of the Laplace operator~\eqref{eq:laplace} with eigenvalues $-l(l+1)/R^2$, we can rewrite the integrals over the quadratic terms in~\eqref{eqn1} as
\begin{align}
\int\dif\Omega\,\phi^2 & =  \sum_{l,m}  \abs{w_{l,m}}^2  , \label{eq:SHintegral}\\
\int\dif\Omega\,\lvert\nabla\phi\rvert^2 & = \sum_{l,m}\frac{l(l+1)}{R^2}\abs{w_{l,m}}^2 ,\\
\int\dif\Omega\,[\bigtriangledown^2\phi]^2 & = \sum_{l,m}\frac{[l(l+1)]^2}{R^4}\abs{w_{l,m}}^2 .
\end{align}

To evaluate the quartic term we use~\eqref{eq:SHintegral} once more, but with $\phi$ replaced by $\phi^2$:
\begin{equation}
\int\dif\Omega\,\phi^4 =  \sum_{l,m}  \abs{u_{l,m}}^2  \\
\end{equation}
where $u_{l,m}$ are the coefficients of the spherical harmonic expansion of the the field $\phi^2$. They can be calculated according to~\eqref{eqn3} as
\begin{equation}
u_{l,m} = \int \dif\Omega \, Y_{l,m}^*(\theta,\varphi) \phi(\theta,\varphi)^2 .
\end{equation}
Taking the complex conjugate of this equation and then substituting~\eqref{eqn2} for each power of the real-valued field $\phi$ gives
\begin{equation}
\begin{split}
u_{l,m}^* = & \sum_{l_1,m_2}\sum_{l_2,m_2}w_{l_1,m_1}w_{l_2,m_2} \\
  &{}\times\int\dif\Omega\,Y_{l,m}(\theta,\varphi)Y_{l_1,m_1}(\theta,\varphi)Y_{l_2,m_2}(\theta,\varphi).
\end{split}
\end{equation}

Performing the integral over the triple product of spherical harmonic functions gives rise to the Wigner 3j symbols~\cite{DLMF}:
\begin{multline}
\int\dif\Omega\,Y_{l_1,m_1}(\theta,\varphi)Y_{l_2,m_2}(\theta,\varphi)Y_{l,m}(\theta,\varphi) =\\
\sqrt{\frac{(2l_1\!+\!1)(2l_2\!+\!1)(2l\!+\!1)}{4\pi}}\begin{pmatrix}l_1&l_2&l\\0&0&0\end{pmatrix}\begin{pmatrix}l_1&l_2&l\\m_1&m_2&m\end{pmatrix}.
 \end{multline}

When combined, these equations result in expression~\eqref{eqn4} for the total system energy.

\section{\label{append2}Mean-Field Phase Diagram for the Planar System}

Here we briefly summarize several key results obtained in Ref.~\onlinecite{Shlomovitz14} for the model~\eqref{eqn0} on a planar surface that are relevant for this work. The field $\phi(\vek{r})$ is defined on a square region of side length $L$. If all parameters are positive then the ground state of the system is the uniform field $\phi(\vek{r}) = 0$, which has energy $E_{\text{homog}} = 0$. If $\alpha < 0$ while $\gamma>0$ then the states $\phi(\vek{r}) = \pm \sqrt{-\alpha/b}$  have a lower energy of $E_\text{coex} = - L^2 \alpha^2 / (4b)$. The boundary between the coexistence region and the homogeneous fluid is the line 
\begin{equation}
\alpha=0. \label{eq:homogcoexboundary}
\end{equation}

To estimate the energy of the modulated phase we expand the order parameter field $\phi(\vek{r})$ in a plane wave basis,
\begin{equation}
\phi(\vek{r})=\frac{1}{L^2} \sum_{\vek{k}} \tilde{\phi}(\vek{k}) e^{i\vek{k}\cdot\vek{r}} \label{eq:Fourierexpansion}
\end{equation} 
Expressed in terms of the Fourier coefficients $\tilde{\phi}(\vek{k})$ the energy becomes
\begin{equation}
\begin{split}
E=
&\frac{1}{2L^2}\sum_{\vek{k}}(\alpha+\gamma k^2+\epsilon k^4) \lvert\tilde{\phi}(\vek{k})\rvert^2\\
&+\frac{b}{4L^6}\sum_{\vek{k},\vek{k}',\vek{k}''}\tilde{\phi}(\vek{k})\tilde{\phi}(\vek{k}')\tilde{\phi}(\vek{k}'')\tilde{\phi}(-\vek{k}-\vek{k}'-\vek{k}'') . \label{eq:flatFourierEnergy}
\end{split}
\end{equation}

Analogous to our approach in Section~\ref{subsec:singlemode} we now assume that only a single Fourier mode with wave vector $\vek{k} \neq 0$ and its Hermitian conjugate contribute to the expansion~\eqref{eq:Fourierexpansion}. Under this assumption the energy is
\begin{equation}
E = \frac{1}{L^2}  (\alpha+\gamma k^2+\epsilon k^4) \lvert\tilde{\phi}(\vek{k})\rvert^2
+ \frac{3b}{2L^6}  \lvert\tilde{\phi}(\vek{k})\rvert^4
\end{equation}
Minimizing this expression with respect to both $k$ and $\abs{\tilde{\phi}(\vek{k})}$, we find that for negative $\gamma$  the ground state in this single mode approximation has the wave vector magnitude $\kstar = \sqrt{-\gamma/(2 \epsilon)}$ and energy
\begin{equation}
E_{\text{modulated}} = - \frac{L^2 (\gamma^2 - 4 \alpha \epsilon)^2}{96 b \epsilon^2} .
\end{equation}

The phase boundary between the modulated and the homogeneous fluid in the region $\alpha>0, \gamma<0$ can be obtained by determining when $E_{\text{modulated}}$ is equal to $E_{\text{homog}}$, which yields
\begin{equation}
\alpha = \frac{\gamma^2}{4 \epsilon} . \label{eq:modulatedhomogboundary}
\end{equation}
Similarly, equating $E_{\text{modulated}}$ and $E_\text{coex}$ for  $\alpha< 0, \gamma<0$ gives
\begin{equation}
\alpha=-\frac{\gamma^2}{2\epsilon}\left(1+\sqrt{\frac{3}{2}}\right) \label{eq:modulatedcoexboundary}
\end{equation}
for the boundary between the modulated phase and the coexistence region. The three curves~\eqref{eq:homogcoexboundary}, \eqref{eq:modulatedhomogboundary} and \eqref{eq:modulatedcoexboundary} for the planar system are included in Figure~\ref{fig1} for comparison with the spherical systems' phase diagrams.

\section{\label{append3}Finite Size Effect in Planar Bilayers}

Here we demonstrate that finite size effects can also occur in planar systems, for example the shift of the Lifshitz line that separates the homogeneous fluid from the microemulsion. Starting with the expression~\eqref{eq:flatFourierEnergy} for the energy, we follow the same approach we used to derive~\eqref{eq:SHgaussian} for the spherical system: we assume that we can ignore the quartic term, and consider the case $b=0$. Since the energy is then a quadratic function of the Fourier coefficients we can immediately determine their variances, and with them the static scattering structure factor~\cite{Luo18a}, from the equipartition theorem:
\begin{equation}
\label{a5}
S(k) \equiv \frac{1}{L^2}\langle\lvert\tilde{\phi}(k)\rvert^2\rangle=\frac{k_\text{B} T}{\alpha+\gamma k^2+\epsilon k^4}. 
\end{equation}
This function is monotonically decreasing if all parameters are positive, but it has a peak at non-zero wave vector $\kstar = \sqrt{-\gamma/(2\epsilon)}$ if $\gamma$ is negative. It is notable that the location of the peak corresponds to the dominant wave vector of the modulated phase,  discussed in the previous section. These results suggest that the Lifshitz line lies at $\gamma=0$ as reported previously~\cite{Shlomovitz13, Shlomovitz14}.

This analysis, however, applies only to an infinitely large system. For a finite system of side length $L$ the wave vectors must be of the form $\vek{k} = (2 \pi m/L, 2\pi n/L)$, where $m$ and $n$ are integers. The smallest wave vector supported by such a system is $\kmin =2\pi/L$. If $\kstar$ is smaller than that then it is possible that the peak in the structure factor will not be measurable in the fluctuation spectrum of the accessible wave vectors.

The transition from a homogeneous fluid to a microemulsion in a finite system occurs when $S(\kmin)$ becomes greater than $S(0)$ for the first time. Equating these two properties we find that this transition occurs at 
\begin{equation}
\gamma=-\frac{4\pi^2\epsilon}{L^2} .
\end{equation}
As expected, in the limit of $L \rightarrow \infty$ the Lifshitz line lies at $\gamma = 0$. For small systems, however, the homogeneous fluid phase is stable even for slightly negative $\gamma$ if the system size is insufficient to accommodate the characteristic fluctuations of the microemulsion phase.

\bibliography{paper}

\begin{thebibliography}{53}%
\makeatletter
\providecommand \@ifxundefined [1]{%
 \@ifx{#1\undefined}
}%
\providecommand \@ifnum [1]{%
 \ifnum #1\expandafter \@firstoftwo
 \else \expandafter \@secondoftwo
 \fi
}%
\providecommand \@ifx [1]{%
 \ifx #1\expandafter \@firstoftwo
 \else \expandafter \@secondoftwo
 \fi
}%
\providecommand \natexlab [1]{#1}%
\providecommand \enquote  [1]{``#1''}%
\providecommand \bibnamefont  [1]{#1}%
\providecommand \bibfnamefont [1]{#1}%
\providecommand \citenamefont [1]{#1}%
\providecommand \href@noop [0]{\@secondoftwo}%
\providecommand \href [0]{\begingroup \@sanitize@url \@href}%
\providecommand \@href[1]{\@@startlink{#1}\@@href}%
\providecommand \@@href[1]{\endgroup#1\@@endlink}%
\providecommand \@sanitize@url [0]{\catcode `\\12\catcode `\$12\catcode
  `\&12\catcode `\#12\catcode `\^12\catcode `\_12\catcode `\%12\relax}%
\providecommand \@@startlink[1]{}%
\providecommand \@@endlink[0]{}%
\providecommand \url  [0]{\begingroup\@sanitize@url \@url }%
\providecommand \@url [1]{\endgroup\@href {#1}{\urlprefix }}%
\providecommand \urlprefix  [0]{URL }%
\providecommand \Eprint [0]{\href }%
\providecommand \doibase [0]{http://dx.doi.org/}%
\providecommand \selectlanguage [0]{\@gobble}%
\providecommand \bibinfo  [0]{\@secondoftwo}%
\providecommand \bibfield  [0]{\@secondoftwo}%
\providecommand \translation [1]{[#1]}%
\providecommand \BibitemOpen [0]{}%
\providecommand \bibitemStop [0]{}%
\providecommand \bibitemNoStop [0]{.\EOS\space}%
\providecommand \EOS [0]{\spacefactor3000\relax}%
\providecommand \BibitemShut  [1]{\csname bibitem#1\endcsname}%
\let\auto@bib@innerbib\@empty
\bibitem [{\citenamefont {Veatch}\ and\ \citenamefont
  {Keller}(2003)}]{Veatch03}%
  \BibitemOpen
  \bibfield  {author} {\bibinfo {author} {\bibfnamefont {S.~L.}\ \bibnamefont
  {Veatch}}\ and\ \bibinfo {author} {\bibfnamefont {S.~L.}\ \bibnamefont
  {Keller}},\ }\bibfield  {title} {\enquote {\bibinfo {title} {Separation of
  liquid phases in giant vesicles of ternary mixtures of phospholipids and
  cholesterol},}\ }\href {\doibase 10.1016/S0006-3495(03)74726-2} {\bibfield
  {journal} {\bibinfo  {journal} {Biophys. J.}\ }\textbf {\bibinfo {volume}
  {85}},\ \bibinfo {pages} {3074--3083} (\bibinfo {year} {2003})}\BibitemShut
  {NoStop}%
\bibitem [{\citenamefont {Scherfeld}\ \emph {et~al.}(2003)\citenamefont
  {Scherfeld}, \citenamefont {Kahya},\ and\ \citenamefont
  {Schwille}}]{Scherfeld03}%
  \BibitemOpen
  \bibfield  {author} {\bibinfo {author} {\bibfnamefont {D.}~\bibnamefont
  {Scherfeld}}, \bibinfo {author} {\bibfnamefont {N.}~\bibnamefont {Kahya}}, \
  and\ \bibinfo {author} {\bibfnamefont {P.}~\bibnamefont {Schwille}},\
  }\bibfield  {title} {\enquote {\bibinfo {title} {Lipid dynamics and domain
  formation in model membranes composed of ternary mixtures of unsaturated and
  saturated phosphatidylcholines and cholesterol},}\ }\href {\doibase
  10.1016/S0006-3495(03)74791-2} {\bibfield  {journal} {\bibinfo  {journal}
  {Biophys. J.}\ }\textbf {\bibinfo {volume} {85}},\ \bibinfo {pages}
  {3758--3768} (\bibinfo {year} {2003})}\BibitemShut {NoStop}%
\bibitem [{\citenamefont {Veatch}\ and\ \citenamefont
  {Keller}(2005)}]{Veatch05}%
  \BibitemOpen
  \bibfield  {author} {\bibinfo {author} {\bibfnamefont {S.~L.}\ \bibnamefont
  {Veatch}}\ and\ \bibinfo {author} {\bibfnamefont {S.~L.}\ \bibnamefont
  {Keller}},\ }\bibfield  {title} {\enquote {\bibinfo {title} {Seeing spots:
  Complex phase behavior in simple membranes},}\ }\href {\doibase
  10.1016/j.bbamcr.2005.06.010} {\bibfield  {journal} {\bibinfo  {journal}
  {Biochim. Biophys. Acta -- Molecular Cell Research}\ }\textbf {\bibinfo
  {volume} {1746}},\ \bibinfo {pages} {172--185} (\bibinfo {year}
  {2005})}\BibitemShut {NoStop}%
\bibitem [{\citenamefont {Marsh}(2009)}]{Marsh09}%
  \BibitemOpen
  \bibfield  {author} {\bibinfo {author} {\bibfnamefont {D.}~\bibnamefont
  {Marsh}},\ }\bibfield  {title} {\enquote {\bibinfo {title}
  {Cholesterol-induced fluid membrane domains: A compendium of lipid-raft
  ternary phase diagrams},}\ }\href {\doibase 10.1016/j.bbamem.2009.08.004}
  {\bibfield  {journal} {\bibinfo  {journal} {Biochim. Biophys. Acta --
  Biomembranes}\ }\textbf {\bibinfo {volume} {1788}},\ \bibinfo {pages}
  {2114--2123} (\bibinfo {year} {2009})}\BibitemShut {NoStop}%
\bibitem [{\citenamefont {Silvius}\ \emph {et~al.}(1996)\citenamefont
  {Silvius}, \citenamefont {{del Giudice}},\ and\ \citenamefont
  {Lafleur}}]{Silvius96}%
  \BibitemOpen
  \bibfield  {author} {\bibinfo {author} {\bibfnamefont {J.~R.}\ \bibnamefont
  {Silvius}}, \bibinfo {author} {\bibfnamefont {D.}~\bibnamefont {{del
  Giudice}}}, \ and\ \bibinfo {author} {\bibfnamefont {M.}~\bibnamefont
  {Lafleur}},\ }\bibfield  {title} {\enquote {\bibinfo {title} {Cholesterol at
  different bilayer concentrations can promote or antagonize lateral
  segregation of phospholipids of differing acyl chain length},}\ }\href
  {\doibase 10.1021/bi9615506} {\bibfield  {journal} {\bibinfo  {journal}
  {Biochemistry}\ }\textbf {\bibinfo {volume} {35}},\ \bibinfo {pages}
  {151988--15208} (\bibinfo {year} {1996})}\BibitemShut {NoStop}%
\bibitem [{\citenamefont {Wang}\ and\ \citenamefont {Silvius}(2001)}]{Wang01}%
  \BibitemOpen
  \bibfield  {author} {\bibinfo {author} {\bibfnamefont {T.-Y.}\ \bibnamefont
  {Wang}}\ and\ \bibinfo {author} {\bibfnamefont {J.~R.}\ \bibnamefont
  {Silvius}},\ }\bibfield  {title} {\enquote {\bibinfo {title} {Cholesterol
  does not induce segregation of liquid-ordered domains in bilayers modeling
  the inner leaflet of the plasma membrane},}\ }\href {\doibase
  10.1016/S0006-3495(01)75919-X} {\bibfield  {journal} {\bibinfo  {journal}
  {Biophys. J.}\ }\textbf {\bibinfo {volume} {81}},\ \bibinfo {pages}
  {2762--2773} (\bibinfo {year} {2001})}\BibitemShut {NoStop}%
\bibitem [{\citenamefont {Veatch}\ \emph {et~al.}(2004)\citenamefont {Veatch},
  \citenamefont {Polozov}, \citenamefont {Gawrisch},\ and\ \citenamefont
  {Keller}}]{Veatch04}%
  \BibitemOpen
  \bibfield  {author} {\bibinfo {author} {\bibfnamefont {S.~L.}\ \bibnamefont
  {Veatch}}, \bibinfo {author} {\bibfnamefont {I.~V.}\ \bibnamefont {Polozov}},
  \bibinfo {author} {\bibfnamefont {K.}~\bibnamefont {Gawrisch}}, \ and\
  \bibinfo {author} {\bibfnamefont {S.~L.}\ \bibnamefont {Keller}},\ }\bibfield
   {title} {\enquote {\bibinfo {title} {Liquid domains in vesicles investigated
  by nmr and fluorescence microscopy},}\ }\href {\doibase
  10.1016/S0006-3495(04)74342-8} {\bibfield  {journal} {\bibinfo  {journal}
  {Biophys. J.}\ }\textbf {\bibinfo {volume} {86}},\ \bibinfo {pages}
  {2910--2922} (\bibinfo {year} {2004})}\BibitemShut {NoStop}%
\bibitem [{\citenamefont {Zhao}\ \emph {et~al.}(2007)\citenamefont {Zhao},
  \citenamefont {Wu}, \citenamefont {Heberle}, \citenamefont {Mills},
  \citenamefont {Klawitter}, \citenamefont {Huang}, \citenamefont {Costanza},\
  and\ \citenamefont {Feigenson}}]{Zhao07}%
  \BibitemOpen
  \bibfield  {author} {\bibinfo {author} {\bibfnamefont {J.}~\bibnamefont
  {Zhao}}, \bibinfo {author} {\bibfnamefont {J.}~\bibnamefont {Wu}}, \bibinfo
  {author} {\bibfnamefont {F.~A.}\ \bibnamefont {Heberle}}, \bibinfo {author}
  {\bibfnamefont {T.~T.}\ \bibnamefont {Mills}}, \bibinfo {author}
  {\bibfnamefont {P.}~\bibnamefont {Klawitter}}, \bibinfo {author}
  {\bibfnamefont {G.}~\bibnamefont {Huang}}, \bibinfo {author} {\bibfnamefont
  {G.}~\bibnamefont {Costanza}}, \ and\ \bibinfo {author} {\bibfnamefont
  {G.~W.}\ \bibnamefont {Feigenson}},\ }\bibfield  {title} {\enquote {\bibinfo
  {title} {Phase studies of model biomembranes: Complex behavior of
  {DSPC}/{DOPC}/{C}holesterol},}\ }\href {\doibase
  10.1016/j.bbamem.2007.07.008} {\bibfield  {journal} {\bibinfo  {journal}
  {Biochim. Biophys. Acta -- Biomembranes}\ }\textbf {\bibinfo {volume}
  {1768}},\ \bibinfo {pages} {2764--2776} (\bibinfo {year} {2007})}\BibitemShut
  {NoStop}%
\bibitem [{\citenamefont {Heberle}\ \emph {et~al.}(2010)\citenamefont
  {Heberle}, \citenamefont {Wu}, \citenamefont {Goh}, \citenamefont
  {Petruzielo},\ and\ \citenamefont {Feigenson}}]{Heberle10}%
  \BibitemOpen
  \bibfield  {author} {\bibinfo {author} {\bibfnamefont {F.~A.}\ \bibnamefont
  {Heberle}}, \bibinfo {author} {\bibfnamefont {J.}~\bibnamefont {Wu}},
  \bibinfo {author} {\bibfnamefont {S.~L.}\ \bibnamefont {Goh}}, \bibinfo
  {author} {\bibfnamefont {R.~S.}\ \bibnamefont {Petruzielo}}, \ and\ \bibinfo
  {author} {\bibfnamefont {G.~W.}\ \bibnamefont {Feigenson}},\ }\bibfield
  {title} {\enquote {\bibinfo {title} {Comparison of three ternary lipid
  bilayer mixtures: {FRET} and {ESR} reveal nanodomains},}\ }\href {\doibase
  10.1016/j.bpj.2010.09.064} {\bibfield  {journal} {\bibinfo  {journal}
  {Biophys. J.}\ }\textbf {\bibinfo {volume} {99}},\ \bibinfo {pages}
  {3309--3318} (\bibinfo {year} {2010})}\BibitemShut {NoStop}%
\bibitem [{\citenamefont {Pencer}\ \emph {et~al.}(2005)\citenamefont {Pencer},
  \citenamefont {Nieh}, \citenamefont {Harroun}, \citenamefont {Krueger},
  \citenamefont {Adams},\ and\ \citenamefont {Katsaras}}]{Pencer05}%
  \BibitemOpen
  \bibfield  {author} {\bibinfo {author} {\bibfnamefont {J.}~\bibnamefont
  {Pencer}}, \bibinfo {author} {\bibfnamefont {M.-P.}\ \bibnamefont {Nieh}},
  \bibinfo {author} {\bibfnamefont {T.~A.}\ \bibnamefont {Harroun}}, \bibinfo
  {author} {\bibfnamefont {S.}~\bibnamefont {Krueger}}, \bibinfo {author}
  {\bibfnamefont {C.}~\bibnamefont {Adams}}, \ and\ \bibinfo {author}
  {\bibfnamefont {J.}~\bibnamefont {Katsaras}},\ }\bibfield  {title} {\enquote
  {\bibinfo {title} {Bilayer thickness and thermal response of
  dimyristoylphosphatidylcholine unilamellar vesicles containing cholesterol,
  ergosterol and lanosterol: A small-angle neutron scattering study},}\ }\href
  {\doibase 10.1016/j.bbamem.2005.10.017} {\bibfield  {journal} {\bibinfo
  {journal} {Biophys. J.}\ }\textbf {\bibinfo {volume} {1720}},\ \bibinfo
  {pages} {84--91} (\bibinfo {year} {2005})}\BibitemShut {NoStop}%
\bibitem [{\citenamefont {Pabst}\ \emph {et~al.}(2010)\citenamefont {Pabst},
  \citenamefont {Ku{\v{c}}erka}, \citenamefont {Nieh}, \citenamefont
  {Rheinst{\"a}dter},\ and\ \citenamefont {Katsaras}}]{Pabst10}%
  \BibitemOpen
  \bibfield  {author} {\bibinfo {author} {\bibfnamefont {G.}~\bibnamefont
  {Pabst}}, \bibinfo {author} {\bibfnamefont {N.}~\bibnamefont
  {Ku{\v{c}}erka}}, \bibinfo {author} {\bibfnamefont {M.-P.}\ \bibnamefont
  {Nieh}}, \bibinfo {author} {\bibfnamefont {M.~C.}\ \bibnamefont
  {Rheinst{\"a}dter}}, \ and\ \bibinfo {author} {\bibfnamefont
  {J.}~\bibnamefont {Katsaras}},\ }\bibfield  {title} {\enquote {\bibinfo
  {title} {Applications of neutron and {X}-ray scattering to the study of
  biologically relevant model membranes},}\ }\href {\doibase
  10.1016/j.chemphyslip.2010.03.010} {\bibfield  {journal} {\bibinfo  {journal}
  {Chem. Phys. Lipids}\ }\textbf {\bibinfo {volume} {163}},\ \bibinfo {pages}
  {460--479} (\bibinfo {year} {2010})}\BibitemShut {NoStop}%
\bibitem [{\citenamefont {Kollmitzer}\ \emph {et~al.}(2015)\citenamefont
  {Kollmitzer}, \citenamefont {Heftberger}, \citenamefont {Podgornik},
  \citenamefont {Nagle},\ and\ \citenamefont {Pabst}}]{Kollmitzer15}%
  \BibitemOpen
  \bibfield  {author} {\bibinfo {author} {\bibfnamefont {B.}~\bibnamefont
  {Kollmitzer}}, \bibinfo {author} {\bibfnamefont {P.}~\bibnamefont
  {Heftberger}}, \bibinfo {author} {\bibfnamefont {R.}~\bibnamefont
  {Podgornik}}, \bibinfo {author} {\bibfnamefont {J.~F.}\ \bibnamefont
  {Nagle}}, \ and\ \bibinfo {author} {\bibfnamefont {G.}~\bibnamefont
  {Pabst}},\ }\bibfield  {title} {\enquote {\bibinfo {title} {Bending
  rigidities and interdomain forces in membranes with coexisting lipid
  domains},}\ }\href {\doibase 10.1016/j.bpj.2015.05.003} {\bibfield  {journal}
  {\bibinfo  {journal} {Biophys. J.}\ }\textbf {\bibinfo {volume} {108}},\
  \bibinfo {pages} {2833--2842} (\bibinfo {year} {2015})}\BibitemShut {NoStop}%
\bibitem [{\citenamefont {Risselada}\ and\ \citenamefont
  {Marrink}(2008)}]{Risselada08}%
  \BibitemOpen
  \bibfield  {author} {\bibinfo {author} {\bibfnamefont {H.~J.}\ \bibnamefont
  {Risselada}}\ and\ \bibinfo {author} {\bibfnamefont {S.~J.}\ \bibnamefont
  {Marrink}},\ }\bibfield  {title} {\enquote {\bibinfo {title} {The molecular
  face of lipid rafts in model membranes},}\ }\href {\doibase
  10.1073/pnas.0807527105} {\bibfield  {journal} {\bibinfo  {journal} {Proc.
  Natl. Acad. Sci. USA}\ }\textbf {\bibinfo {volume} {105}},\ \bibinfo {pages}
  {17367--17372} (\bibinfo {year} {2008})}\BibitemShut {NoStop}%
\bibitem [{\citenamefont {Rosetti}\ and\ \citenamefont
  {Pastorino}(2010)}]{Rosetti11}%
  \BibitemOpen
  \bibfield  {author} {\bibinfo {author} {\bibfnamefont {C.}~\bibnamefont
  {Rosetti}}\ and\ \bibinfo {author} {\bibfnamefont {C.}~\bibnamefont
  {Pastorino}},\ }\bibfield  {title} {\enquote {\bibinfo {title}
  {Polyunsaturated and saturated phospholipids in mixed bilayers: A study from
  the molecular scale to the lateral lipid organization},}\ }\href {\doibase
  10.1021/jp1082888} {\bibfield  {journal} {\bibinfo  {journal} {J. Phys. Chem.
  B}\ }\textbf {\bibinfo {volume} {115}},\ \bibinfo {pages} {1002--1013}
  (\bibinfo {year} {2010})}\BibitemShut {NoStop}%
\bibitem [{\citenamefont {Sodt}\ \emph {et~al.}(2014)\citenamefont {Sodt},
  \citenamefont {Sandar}, \citenamefont {Gawrisch}, \citenamefont {Pastor},\
  and\ \citenamefont {Lyman}}]{Sodt14}%
  \BibitemOpen
  \bibfield  {author} {\bibinfo {author} {\bibfnamefont {A.~J.}\ \bibnamefont
  {Sodt}}, \bibinfo {author} {\bibfnamefont {M.~L.}\ \bibnamefont {Sandar}},
  \bibinfo {author} {\bibfnamefont {K.}~\bibnamefont {Gawrisch}}, \bibinfo
  {author} {\bibfnamefont {R.~W.}\ \bibnamefont {Pastor}}, \ and\ \bibinfo
  {author} {\bibfnamefont {E.}~\bibnamefont {Lyman}},\ }\bibfield  {title}
  {\enquote {\bibinfo {title} {The molecular structure of the liquid-ordered
  phase of lipid bilayers},}\ }\href {\doibase 10.1021/ja4105667} {\bibfield
  {journal} {\bibinfo  {journal} {J. Am. Chem. Soc.}\ }\textbf {\bibinfo
  {volume} {136}},\ \bibinfo {pages} {725--732} (\bibinfo {year}
  {2014})}\BibitemShut {NoStop}%
\bibitem [{\citenamefont {Baoukina}\ \emph {et~al.}(2017)\citenamefont
  {Baoukina}, \citenamefont {Rozmanov},\ and\ \citenamefont
  {Tieleman}}]{Baoukina17}%
  \BibitemOpen
  \bibfield  {author} {\bibinfo {author} {\bibfnamefont {S.}~\bibnamefont
  {Baoukina}}, \bibinfo {author} {\bibfnamefont {D.}~\bibnamefont {Rozmanov}},
  \ and\ \bibinfo {author} {\bibfnamefont {D.~P.}\ \bibnamefont {Tieleman}},\
  }\bibfield  {title} {\enquote {\bibinfo {title} {Composition fluctuations in
  lipid bilayers},}\ }\href {\doibase 10.1016/j.bpj.2017.10.009} {\bibfield
  {journal} {\bibinfo  {journal} {Biophys. J.}\ }\textbf {\bibinfo {volume}
  {113}},\ \bibinfo {pages} {2750--2761} (\bibinfo {year} {2017})}\BibitemShut
  {NoStop}%
\bibitem [{\citenamefont {Pantelopulos}\ \emph {et~al.}(2017)\citenamefont
  {Pantelopulos}, \citenamefont {Nagai}, \citenamefont {Bandara}, \citenamefont
  {Panahi},\ and\ \citenamefont {Straub}}]{Pantelopulos17}%
  \BibitemOpen
  \bibfield  {author} {\bibinfo {author} {\bibfnamefont {G.~A.}\ \bibnamefont
  {Pantelopulos}}, \bibinfo {author} {\bibfnamefont {T.}~\bibnamefont {Nagai}},
  \bibinfo {author} {\bibfnamefont {A.}~\bibnamefont {Bandara}}, \bibinfo
  {author} {\bibfnamefont {A.}~\bibnamefont {Panahi}}, \ and\ \bibinfo {author}
  {\bibfnamefont {J.~E.}\ \bibnamefont {Straub}},\ }\bibfield  {title}
  {\enquote {\bibinfo {title} {Critical size dependence of domain formation
  observed in coarse-grained simulations of bilayers composed of ternary lipid
  mixtures},}\ }\href {\doibase 10.1063/1.4999709} {\bibfield  {journal}
  {\bibinfo  {journal} {J. Chem. Phys.}\ }\textbf {\bibinfo {volume} {147}},\
  \bibinfo {pages} {095101} (\bibinfo {year} {2017})}\BibitemShut {NoStop}%
\bibitem [{\citenamefont {Radhakrishnan}\ and\ \citenamefont
  {McConnell}(2005)}]{Radhakrishnan05}%
  \BibitemOpen
  \bibfield  {author} {\bibinfo {author} {\bibfnamefont {A.}~\bibnamefont
  {Radhakrishnan}}\ and\ \bibinfo {author} {\bibfnamefont {H.~M.}\ \bibnamefont
  {McConnell}},\ }\bibfield  {title} {\enquote {\bibinfo {title} {Condensed
  complexes in vesicles containing cholesterol and phospholipids},}\ }\href
  {\doibase 10.1073/pnas.0506043102} {\bibfield  {journal} {\bibinfo  {journal}
  {Proc. Natl. Acad. Sci. USA}\ }\textbf {\bibinfo {volume} {102}},\ \bibinfo
  {pages} {12662--12666} (\bibinfo {year} {2005})}\BibitemShut {NoStop}%
\bibitem [{\citenamefont {Almeida}(2009)}]{Almeida09}%
  \BibitemOpen
  \bibfield  {author} {\bibinfo {author} {\bibfnamefont {P.~F.~F.}\
  \bibnamefont {Almeida}},\ }\bibfield  {title} {\enquote {\bibinfo {title}
  {Thermodynamics of lipid interactions in complex bilayers},}\ }\href
  {\doibase 10.1016/j.bbamem.2008.08.007} {\bibfield  {journal} {\bibinfo
  {journal} {Biochim. Biophys. Acta -- Biomembranes}\ }\textbf {\bibinfo
  {volume} {1788}},\ \bibinfo {pages} {72--85} (\bibinfo {year}
  {2009})}\BibitemShut {NoStop}%
\bibitem [{\citenamefont {Putzel}\ and\ \citenamefont
  {Schick}(2011)}]{Putzel11b}%
  \BibitemOpen
  \bibfield  {author} {\bibinfo {author} {\bibfnamefont {G.~G.}\ \bibnamefont
  {Putzel}}\ and\ \bibinfo {author} {\bibfnamefont {M.}~\bibnamefont
  {Schick}},\ }\bibfield  {title} {\enquote {\bibinfo {title} {Insights on raft
  behavior from minimal phenomenological models},}\ }\href {\doibase
  10.1088/0953-8984/23/28/284101} {\bibfield  {journal} {\bibinfo  {journal}
  {J. Phys. Cond. Mat.}\ }\textbf {\bibinfo {volume} {23}},\ \bibinfo {pages}
  {284101} (\bibinfo {year} {2011})}\BibitemShut {NoStop}%
\bibitem [{\citenamefont {Svetlovics}\ \emph {et~al.}(2012)\citenamefont
  {Svetlovics}, \citenamefont {Wheaten},\ and\ \citenamefont
  {Almeida}}]{Svetlovics12}%
  \BibitemOpen
  \bibfield  {author} {\bibinfo {author} {\bibfnamefont {J.~A.}\ \bibnamefont
  {Svetlovics}}, \bibinfo {author} {\bibfnamefont {S.~A.}\ \bibnamefont
  {Wheaten}}, \ and\ \bibinfo {author} {\bibfnamefont {P.~F.}\ \bibnamefont
  {Almeida}},\ }\bibfield  {title} {\enquote {\bibinfo {title} {Phase
  separation and fluctuations in mixtures of a saturated and an unsaturated
  phospholipid},}\ }\href {\doibase 10.1016/j.bpj.2012.04.017} {\bibfield
  {journal} {\bibinfo  {journal} {Biophys. J.}\ }\textbf {\bibinfo {volume}
  {102}},\ \bibinfo {pages} {2526--2535} (\bibinfo {year} {2012})}\BibitemShut
  {NoStop}%
\bibitem [{\citenamefont {Simons}\ and\ \citenamefont
  {Ikonen}(1997)}]{Simons97}%
  \BibitemOpen
  \bibfield  {author} {\bibinfo {author} {\bibfnamefont {K.}~\bibnamefont
  {Simons}}\ and\ \bibinfo {author} {\bibfnamefont {E.}~\bibnamefont
  {Ikonen}},\ }\bibfield  {title} {\enquote {\bibinfo {title} {Functional rafts
  in cell membranes},}\ }\href {\doibase 10.1038/42408} {\bibfield  {journal}
  {\bibinfo  {journal} {Nature}\ }\textbf {\bibinfo {volume} {387}},\ \bibinfo
  {pages} {569--572} (\bibinfo {year} {1997})}\BibitemShut {NoStop}%
\bibitem [{\citenamefont {Brown}\ and\ \citenamefont {London}(1998)}]{Brown98}%
  \BibitemOpen
  \bibfield  {author} {\bibinfo {author} {\bibfnamefont {D.~A.}\ \bibnamefont
  {Brown}}\ and\ \bibinfo {author} {\bibfnamefont {E.}~\bibnamefont {London}},\
  }\bibfield  {title} {\enquote {\bibinfo {title} {Functions of lipid rafts in
  biological membranes},}\ }\href {\doibase 10.1146/annurev.cellbio.14.1.111}
  {\bibfield  {journal} {\bibinfo  {journal} {Ann. Rev. Cell Dev. Biol.}\
  }\textbf {\bibinfo {volume} {14}},\ \bibinfo {pages} {111--136} (\bibinfo
  {year} {1998})}\BibitemShut {NoStop}%
\bibitem [{\citenamefont {Lingwood}\ and\ \citenamefont
  {Simons}(2010)}]{Lingwood10}%
  \BibitemOpen
  \bibfield  {author} {\bibinfo {author} {\bibfnamefont {D.}~\bibnamefont
  {Lingwood}}\ and\ \bibinfo {author} {\bibfnamefont {K.}~\bibnamefont
  {Simons}},\ }\bibfield  {title} {\enquote {\bibinfo {title} {Lipid rafts as a
  membrane-organizing principle},}\ }\href {\doibase 10.1126/science.1174621}
  {\bibfield  {journal} {\bibinfo  {journal} {Science}\ }\textbf {\bibinfo
  {volume} {327}},\ \bibinfo {pages} {46--50} (\bibinfo {year}
  {2010})}\BibitemShut {NoStop}%
\bibitem [{\citenamefont {Baumgart}\ \emph {et~al.}(2007)\citenamefont
  {Baumgart}, \citenamefont {Hammond}, \citenamefont {Sengupta}, \citenamefont
  {Hess}, \citenamefont {Holowka}, \citenamefont {Baird},\ and\ \citenamefont
  {Webb}}]{Baumgart07}%
  \BibitemOpen
  \bibfield  {author} {\bibinfo {author} {\bibfnamefont {T.}~\bibnamefont
  {Baumgart}}, \bibinfo {author} {\bibfnamefont {A.~T.}\ \bibnamefont
  {Hammond}}, \bibinfo {author} {\bibfnamefont {P.}~\bibnamefont {Sengupta}},
  \bibinfo {author} {\bibfnamefont {S.~T.}\ \bibnamefont {Hess}}, \bibinfo
  {author} {\bibfnamefont {D.~A.}\ \bibnamefont {Holowka}}, \bibinfo {author}
  {\bibfnamefont {B.~A.}\ \bibnamefont {Baird}}, \ and\ \bibinfo {author}
  {\bibfnamefont {W.~W.}\ \bibnamefont {Webb}},\ }\bibfield  {title} {\enquote
  {\bibinfo {title} {Large-scale fluid/fluid phase separation of proteins and
  lipids in giant plasma membrane vesicles},}\ }\href {\doibase
  10.1073/pnas.0611357104} {\bibfield  {journal} {\bibinfo  {journal} {Proc.
  Natl. Acad. Sci. USA}\ }\textbf {\bibinfo {volume} {104}},\ \bibinfo {pages}
  {3165--3170} (\bibinfo {year} {2007})}\BibitemShut {NoStop}%
\bibitem [{\citenamefont {Veatch}\ \emph {et~al.}(2008)\citenamefont {Veatch},
  \citenamefont {Cicuta}, \citenamefont {Sengupta}, \citenamefont
  {Honerkamp-{S}mith}, \citenamefont {Holowka},\ and\ \citenamefont
  {Baird}}]{Veatch08}%
  \BibitemOpen
  \bibfield  {author} {\bibinfo {author} {\bibfnamefont {S.~L.}\ \bibnamefont
  {Veatch}}, \bibinfo {author} {\bibfnamefont {P.}~\bibnamefont {Cicuta}},
  \bibinfo {author} {\bibfnamefont {P.}~\bibnamefont {Sengupta}}, \bibinfo
  {author} {\bibfnamefont {A.}~\bibnamefont {Honerkamp-{S}mith}}, \bibinfo
  {author} {\bibfnamefont {D.}~\bibnamefont {Holowka}}, \ and\ \bibinfo
  {author} {\bibfnamefont {B.}~\bibnamefont {Baird}},\ }\bibfield  {title}
  {\enquote {\bibinfo {title} {Critical fluctuations in plasma membrane
  vesicles},}\ }\href {\doibase 10.1021/cb800012x} {\bibfield  {journal}
  {\bibinfo  {journal} {Chem. Biol.}\ }\textbf {\bibinfo {volume} {3}},\
  \bibinfo {pages} {287--293} (\bibinfo {year} {2008})}\BibitemShut {NoStop}%
\bibitem [{\citenamefont {Levental}\ and\ \citenamefont
  {Levental}(2015)}]{Levental15}%
  \BibitemOpen
  \bibfield  {author} {\bibinfo {author} {\bibfnamefont {K.~R.}\ \bibnamefont
  {Levental}}\ and\ \bibinfo {author} {\bibfnamefont {I.}~\bibnamefont
  {Levental}},\ }\bibfield  {title} {\enquote {\bibinfo {title} {Giant plasma
  membrane vesicles: Models for understanding membrane organization},}\ }\href
  {\doibase 10.1016/bs.ctm.2015.03.009} {\bibfield  {journal} {\bibinfo
  {journal} {Current Topics in Membranes}\ }\textbf {\bibinfo {volume} {75}},\
  \bibinfo {pages} {25--57} (\bibinfo {year} {2015})}\BibitemShut {NoStop}%
\bibitem [{\citenamefont {Konyakhina}\ \emph {et~al.}(2011)\citenamefont
  {Konyakhina}, \citenamefont {Goh}, \citenamefont {Amazon}, \citenamefont
  {Heberle}, \citenamefont {Wu},\ and\ \citenamefont
  {Feigenson}}]{Konyakhina11}%
  \BibitemOpen
  \bibfield  {author} {\bibinfo {author} {\bibfnamefont {T.~M.}\ \bibnamefont
  {Konyakhina}}, \bibinfo {author} {\bibfnamefont {S.~L.}\ \bibnamefont {Goh}},
  \bibinfo {author} {\bibfnamefont {J.}~\bibnamefont {Amazon}}, \bibinfo
  {author} {\bibfnamefont {F.~A.}\ \bibnamefont {Heberle}}, \bibinfo {author}
  {\bibfnamefont {J.}~\bibnamefont {Wu}}, \ and\ \bibinfo {author}
  {\bibfnamefont {G.~W.}\ \bibnamefont {Feigenson}},\ }\bibfield  {title}
  {\enquote {\bibinfo {title} {Control of a nanoscopic-to-macroscopic
  transition: {M}odulated phases in four-component {DSPC}/{DOPC}/{POPC}/{C}hol
  giant unilamellar vesicles},}\ }\href {\doibase 10.1016/j.bpj.2011.06.019}
  {\bibfield  {journal} {\bibinfo  {journal} {Biophys. J.}\ }\textbf {\bibinfo
  {volume} {101}},\ \bibinfo {pages} {L8--L10} (\bibinfo {year}
  {2011})}\BibitemShut {NoStop}%
\bibitem [{\citenamefont {Goh}\ \emph {et~al.}(2013)\citenamefont {Goh},
  \citenamefont {Amazon},\ and\ \citenamefont {Feigenson}}]{Goh13}%
  \BibitemOpen
  \bibfield  {author} {\bibinfo {author} {\bibfnamefont {S.~L.}\ \bibnamefont
  {Goh}}, \bibinfo {author} {\bibfnamefont {J.~J.}\ \bibnamefont {Amazon}}, \
  and\ \bibinfo {author} {\bibfnamefont {G.~W.}\ \bibnamefont {Feigenson}},\
  }\bibfield  {title} {\enquote {\bibinfo {title} {Toward a better raft model:
  Modulated phases in the four-component bilayer,
  {DSPC}/{DOPC}/{POPC}/{C}hol},}\ }\href {\doibase 10.1016/j.bpj.2013.01.003}
  {\bibfield  {journal} {\bibinfo  {journal} {Biophys. J.}\ }\textbf {\bibinfo
  {volume} {104}},\ \bibinfo {pages} {853--862} (\bibinfo {year}
  {2013})}\BibitemShut {NoStop}%
\bibitem [{\citenamefont {Heberle}\ \emph {et~al.}(2013)\citenamefont
  {Heberle}, \citenamefont {Petruzielo}, \citenamefont {Pan}, \citenamefont
  {Drazba}, \citenamefont {Ku{\v{c}}erka}, \citenamefont {Standaert},
  \citenamefont {Feigenson},\ and\ \citenamefont {Katsaras}}]{Heberle13b}%
  \BibitemOpen
  \bibfield  {author} {\bibinfo {author} {\bibfnamefont {F.~A.}\ \bibnamefont
  {Heberle}}, \bibinfo {author} {\bibfnamefont {R.~S.}\ \bibnamefont
  {Petruzielo}}, \bibinfo {author} {\bibfnamefont {J.}~\bibnamefont {Pan}},
  \bibinfo {author} {\bibfnamefont {P.}~\bibnamefont {Drazba}}, \bibinfo
  {author} {\bibfnamefont {N.}~\bibnamefont {Ku{\v{c}}erka}}, \bibinfo {author}
  {\bibfnamefont {R.~F.}\ \bibnamefont {Standaert}}, \bibinfo {author}
  {\bibfnamefont {G.~W.}\ \bibnamefont {Feigenson}}, \ and\ \bibinfo {author}
  {\bibfnamefont {J.}~\bibnamefont {Katsaras}},\ }\bibfield  {title} {\enquote
  {\bibinfo {title} {Bilayer thickness mismatch controls domain size in model
  membranes},}\ }\href {\doibase 10.1021/ja3113615} {\bibfield  {journal}
  {\bibinfo  {journal} {J. Am. Chem. Soc.}\ }\textbf {\bibinfo {volume}
  {135}},\ \bibinfo {pages} {6853--6859} (\bibinfo {year} {2013})}\BibitemShut
  {NoStop}%
\bibitem [{\citenamefont {Konyakhina}\ \emph {et~al.}(2013)\citenamefont
  {Konyakhina}, \citenamefont {Wu}, \citenamefont {Mastroianni}, \citenamefont
  {Heberle},\ and\ \citenamefont {Feigenson}}]{Konyakhina13}%
  \BibitemOpen
  \bibfield  {author} {\bibinfo {author} {\bibfnamefont {T.~M.}\ \bibnamefont
  {Konyakhina}}, \bibinfo {author} {\bibfnamefont {J.}~\bibnamefont {Wu}},
  \bibinfo {author} {\bibfnamefont {J.~D.}\ \bibnamefont {Mastroianni}},
  \bibinfo {author} {\bibfnamefont {F.~A.}\ \bibnamefont {Heberle}}, \ and\
  \bibinfo {author} {\bibfnamefont {G.~W.}\ \bibnamefont {Feigenson}},\
  }\bibfield  {title} {\enquote {\bibinfo {title} {Phase diagram of a
  4-component lipid mixture: {DSPC}/{DOPC}/{POPC}/chol},}\ }\href {\doibase
  10.1016/j.bbamem.2013.05.020} {\bibfield  {journal} {\bibinfo  {journal}
  {Biochim. Biophys. Acta -- Biomembranes}\ }\textbf {\bibinfo {volume}
  {1828}},\ \bibinfo {pages} {2204--2214} (\bibinfo {year} {2013})}\BibitemShut
  {NoStop}%
\bibitem [{\citenamefont {Ackerman}\ and\ \citenamefont
  {Feigenson}(2015)}]{Ackerman15}%
  \BibitemOpen
  \bibfield  {author} {\bibinfo {author} {\bibfnamefont {D.~G.}\ \bibnamefont
  {Ackerman}}\ and\ \bibinfo {author} {\bibfnamefont {G.~W.}\ \bibnamefont
  {Feigenson}},\ }\bibfield  {title} {\enquote {\bibinfo {title} {Multiscale
  modeling of four-component lipid mixtures: Domain composition, size,
  alignment, and properties of the phase interface},}\ }\href {\doibase
  10.1021/jp511083z} {\bibfield  {journal} {\bibinfo  {journal} {J. Phys. Chem.
  B}\ }\textbf {\bibinfo {volume} {119}},\ \bibinfo {pages} {4240--4250}
  (\bibinfo {year} {2015})}\BibitemShut {NoStop}%
\bibitem [{\citenamefont {He}\ and\ \citenamefont {Maibaum}(2018)}]{He18}%
  \BibitemOpen
  \bibfield  {author} {\bibinfo {author} {\bibfnamefont {S.}~\bibnamefont
  {He}}\ and\ \bibinfo {author} {\bibfnamefont {L.}~\bibnamefont {Maibaum}},\
  }\bibfield  {title} {\enquote {\bibinfo {title} {Identifying the onset of
  phase separation in quaternary lipid bilayer systems from coarse-grained
  simulations},}\ }\href {\doibase 10.1021/acs.jpcb.8b00364} {\bibfield
  {journal} {\bibinfo  {journal} {J. Phys. Chem. B}\ }\textbf {\bibinfo
  {volume} {122}},\ \bibinfo {pages} {3961--3973} (\bibinfo {year}
  {2018})}\BibitemShut {NoStop}%
\bibitem [{\citenamefont {Amazon}\ \emph {et~al.}(2013)\citenamefont {Amazon},
  \citenamefont {Goh},\ and\ \citenamefont {Feigenson}}]{Amazon13}%
  \BibitemOpen
  \bibfield  {author} {\bibinfo {author} {\bibfnamefont {J.~J.}\ \bibnamefont
  {Amazon}}, \bibinfo {author} {\bibfnamefont {S.~L.}\ \bibnamefont {Goh}}, \
  and\ \bibinfo {author} {\bibfnamefont {G.~W.}\ \bibnamefont {Feigenson}},\
  }\bibfield  {title} {\enquote {\bibinfo {title} {Competition between line
  tension and curvature stabilizes modulated phase patterns on the surface of
  giant unilamellar vesicles: A simulation study},}\ }\href {\doibase
  10.1103/PhysRevE.87.022708} {\bibfield  {journal} {\bibinfo  {journal} {Phys.
  Rev. E}\ }\textbf {\bibinfo {volume} {87}},\ \bibinfo {pages} {022708}
  (\bibinfo {year} {2013})}\BibitemShut {NoStop}%
\bibitem [{\citenamefont {Shlomovitz}\ and\ \citenamefont
  {Schick}(2013)}]{Shlomovitz13}%
  \BibitemOpen
  \bibfield  {author} {\bibinfo {author} {\bibfnamefont {R.}~\bibnamefont
  {Shlomovitz}}\ and\ \bibinfo {author} {\bibfnamefont {M.}~\bibnamefont
  {Schick}},\ }\bibfield  {title} {\enquote {\bibinfo {title} {Model of a raft
  in both leaves of an asymmetric lipid bilayer},}\ }\href {\doibase
  10.1016/j.bpj.2013.06.053} {\bibfield  {journal} {\bibinfo  {journal}
  {Biophys. J.}\ }\textbf {\bibinfo {volume} {105}},\ \bibinfo {pages}
  {1406--1413} (\bibinfo {year} {2013})}\BibitemShut {NoStop}%
\bibitem [{\citenamefont {Palmieri}\ and\ \citenamefont
  {Safran}(2013)}]{Palmieri13}%
  \BibitemOpen
  \bibfield  {author} {\bibinfo {author} {\bibfnamefont {B.}~\bibnamefont
  {Palmieri}}\ and\ \bibinfo {author} {\bibfnamefont {S.~A.}\ \bibnamefont
  {Safran}},\ }\bibfield  {title} {\enquote {\bibinfo {title} {Hybrid lipids
  increase the probability of fluctuating nanodomains in mixed membranes},}\
  }\href {\doibase 10.1021/la4006168} {\bibfield  {journal} {\bibinfo
  {journal} {Langmuir}\ }\textbf {\bibinfo {volume} {29}},\ \bibinfo {pages}
  {5246--5261} (\bibinfo {year} {2013})}\BibitemShut {NoStop}%
\bibitem [{\citenamefont {Goldenfeld}(1992)}]{Goldenfeld}%
  \BibitemOpen
  \bibfield  {author} {\bibinfo {author} {\bibfnamefont {N.}~\bibnamefont
  {Goldenfeld}},\ }\href@noop {} {\emph {\bibinfo {title} {Lectures on Phase
  Transitions and the Renormalization Group}}}\ (\bibinfo  {publisher} {Perseus
  Books Publishing},\ \bibinfo {address} {Reading, Massachusetts},\ \bibinfo
  {year} {1992})\BibitemShut {NoStop}%
\bibitem [{\citenamefont {Safran}(2003)}]{Safran03}%
  \BibitemOpen
  \bibfield  {author} {\bibinfo {author} {\bibfnamefont {S.~A.}\ \bibnamefont
  {Safran}},\ }\href@noop {} {\emph {\bibinfo {title} {Statistical
  Thermodynamics of Surfaces, Interfaces and Membranes}}},\ Frontiers in
  Physics\ (\bibinfo  {publisher} {Westview Press},\ \bibinfo {address}
  {Boulder, Colorado},\ \bibinfo {year} {2003})\BibitemShut {NoStop}%
\bibitem [{\citenamefont {Shlomovitz}\ \emph {et~al.}(2014)\citenamefont
  {Shlomovitz}, \citenamefont {Maibaum},\ and\ \citenamefont
  {Schick}}]{Shlomovitz14}%
  \BibitemOpen
  \bibfield  {author} {\bibinfo {author} {\bibfnamefont {R.}~\bibnamefont
  {Shlomovitz}}, \bibinfo {author} {\bibfnamefont {L.}~\bibnamefont {Maibaum}},
  \ and\ \bibinfo {author} {\bibfnamefont {M.}~\bibnamefont {Schick}},\
  }\bibfield  {title} {\enquote {\bibinfo {title} {Macroscopic phase
  separation, modulated phases, and microemulsions: A unified picture of
  rafts},}\ }\href {\doibase 10.1016/j.bpj.2014.03.017} {\bibfield  {journal}
  {\bibinfo  {journal} {Biophys. J.}\ }\textbf {\bibinfo {volume} {106}},\
  \bibinfo {pages} {1979--1985} (\bibinfo {year} {2014})}\BibitemShut {NoStop}%
\bibitem [{\citenamefont {Luo}\ and\ \citenamefont {Maibaum}(2018)}]{Luo18a}%
  \BibitemOpen
  \bibfield  {author} {\bibinfo {author} {\bibfnamefont {Y.}~\bibnamefont
  {Luo}}\ and\ \bibinfo {author} {\bibfnamefont {L.}~\bibnamefont {Maibaum}},\
  }\bibfield  {title} {\enquote {\bibinfo {title} {Relating the structure
  factors of two-dimensional materials in planar and spherical geometries},}\
  }\href {\doibase DOI:10.1039/c8sm00978c} {\bibfield  {journal} {\bibinfo
  {journal} {Soft Matter}\ } (\bibinfo {year} {2018}),\
  DOI:10.1039/c8sm00978c}\BibitemShut {NoStop}%
\bibitem [{\citenamefont {Lavrentovich}\ \emph {et~al.}(2016)\citenamefont
  {Lavrentovich}, \citenamefont {Horsley}, \citenamefont {Radja}, \citenamefont
  {Sweeney},\ and\ \citenamefont {Kamien}}]{Lavrentovich16}%
  \BibitemOpen
  \bibfield  {author} {\bibinfo {author} {\bibfnamefont {M.~O.}\ \bibnamefont
  {Lavrentovich}}, \bibinfo {author} {\bibfnamefont {E.~M.}\ \bibnamefont
  {Horsley}}, \bibinfo {author} {\bibfnamefont {A.}~\bibnamefont {Radja}},
  \bibinfo {author} {\bibfnamefont {A.~M.}\ \bibnamefont {Sweeney}}, \ and\
  \bibinfo {author} {\bibfnamefont {R.~D.}\ \bibnamefont {Kamien}},\ }\bibfield
   {title} {\enquote {\bibinfo {title} {First-order patterning transitions on a
  sphere as a route to cell morphology},}\ }\href {\doibase
  10.1073/pnas.1600296113} {\bibfield  {journal} {\bibinfo  {journal} {Proc.
  Natl. Acad. Sci. USA}\ }\textbf {\bibinfo {volume} {113}},\ \bibinfo {pages}
  {5189--5194} (\bibinfo {year} {2016})}\BibitemShut {NoStop}%
\bibitem [{\citenamefont {Schick}(2012)}]{Schick12}%
  \BibitemOpen
  \bibfield  {author} {\bibinfo {author} {\bibfnamefont {M.}~\bibnamefont
  {Schick}},\ }\bibfield  {title} {\enquote {\bibinfo {title} {Membrane
  heterogeneity: Manifestation of a curvature-induced microemulsion},}\ }\href
  {\doibase 10.1103/PhysRevE.85.031902} {\bibfield  {journal} {\bibinfo
  {journal} {Phys. Rev. E}\ }\textbf {\bibinfo {volume} {85}},\ \bibinfo
  {pages} {031902} (\bibinfo {year} {2012})}\BibitemShut {NoStop}%
\bibitem [{\citenamefont {Wieczorek}\ \emph {et~al.}(2016)\citenamefont
  {Wieczorek}, \citenamefont {Meschede}, \citenamefont {Oshchepkov},\ and\
  \citenamefont {{Sales de Andrade}}}]{SHTOOLS34}%
  \BibitemOpen
  \bibfield  {author} {\bibinfo {author} {\bibfnamefont {M.~A.}\ \bibnamefont
  {Wieczorek}}, \bibinfo {author} {\bibfnamefont {M.}~\bibnamefont {Meschede}},
  \bibinfo {author} {\bibfnamefont {I.}~\bibnamefont {Oshchepkov}}, \ and\
  \bibinfo {author} {\bibfnamefont {E.}~\bibnamefont {{Sales de Andrade}}},\
  }\href {\doibase 10.5281/zenodo.592762} {\enquote {\bibinfo {title}
  {{SHTOOLS} library version 3.4},}\ } (\bibinfo {year} {2016})\BibitemShut
  {NoStop}%
\bibitem [{\citenamefont {Sapp}\ \emph {et~al.}(2014)\citenamefont {Sapp},
  \citenamefont {Shlomovitz},\ and\ \citenamefont {Maibaum}}]{Sapp14}%
  \BibitemOpen
  \bibfield  {author} {\bibinfo {author} {\bibfnamefont {K.}~\bibnamefont
  {Sapp}}, \bibinfo {author} {\bibfnamefont {R.}~\bibnamefont {Shlomovitz}}, \
  and\ \bibinfo {author} {\bibfnamefont {L.}~\bibnamefont {Maibaum}},\
  }\bibfield  {title} {\enquote {\bibinfo {title} {Seeing the forest in lieu of
  the trees: Continuum simulations of cell membranes at large length scales},}\
  }in\ \href {\doibase 10.1016/B978-0-444-63378-1.00003-3} {\emph {\bibinfo
  {booktitle} {Annual Reports in Computational Chemistry}}},\ Vol.~\bibinfo
  {volume} {10},\ \bibinfo {editor} {edited by\ \bibinfo {editor}
  {\bibfnamefont {Ralph~A.}\ \bibnamefont {Wheeler}}}\ (\bibinfo  {publisher}
  {Elsevier},\ \bibinfo {address} {Amsterdam},\ \bibinfo {year} {2014})\ pp.\
  \bibinfo {pages} {47--76}\BibitemShut {NoStop}%
\bibitem [{\citenamefont {Baumgart}\ \emph {et~al.}(2003)\citenamefont
  {Baumgart}, \citenamefont {Hess},\ and\ \citenamefont {Webb}}]{Baumgart03}%
  \BibitemOpen
  \bibfield  {author} {\bibinfo {author} {\bibfnamefont {T.}~\bibnamefont
  {Baumgart}}, \bibinfo {author} {\bibfnamefont {S.~T.}\ \bibnamefont {Hess}},
  \ and\ \bibinfo {author} {\bibfnamefont {W.~W.}\ \bibnamefont {Webb}},\
  }\bibfield  {title} {\enquote {\bibinfo {title} {Imaging coexisting fluid
  domains in biomembrane models coupling curvature and line tension},}\ }\href
  {\doibase 10.1038/nature02013} {\bibfield  {journal} {\bibinfo  {journal}
  {Nature}\ }\textbf {\bibinfo {volume} {425}},\ \bibinfo {pages} {821--824}
  (\bibinfo {year} {2003})}\BibitemShut {NoStop}%
\bibitem [{\citenamefont {Nickels}\ \emph {et~al.}(2015)\citenamefont
  {Nickels}, \citenamefont {Cheng}, \citenamefont {Mostofian}, \citenamefont
  {Stanley}, \citenamefont {Lindner}, \citenamefont {Heberle}, \citenamefont
  {Perticaroli}, \citenamefont {Feygenson}, \citenamefont {Egami},
  \citenamefont {Standaert}, \citenamefont {Smith}, \citenamefont {Myles},
  \citenamefont {Ohl},\ and\ \citenamefont {Katsaras}}]{Nickels15}%
  \BibitemOpen
  \bibfield  {author} {\bibinfo {author} {\bibfnamefont {J.~D.}\ \bibnamefont
  {Nickels}}, \bibinfo {author} {\bibfnamefont {X.}~\bibnamefont {Cheng}},
  \bibinfo {author} {\bibfnamefont {B.}~\bibnamefont {Mostofian}}, \bibinfo
  {author} {\bibfnamefont {C.}~\bibnamefont {Stanley}}, \bibinfo {author}
  {\bibfnamefont {B.}~\bibnamefont {Lindner}}, \bibinfo {author} {\bibfnamefont
  {F.~A.}\ \bibnamefont {Heberle}}, \bibinfo {author} {\bibfnamefont
  {S.}~\bibnamefont {Perticaroli}}, \bibinfo {author} {\bibfnamefont
  {M.}~\bibnamefont {Feygenson}}, \bibinfo {author} {\bibfnamefont
  {T.}~\bibnamefont {Egami}}, \bibinfo {author} {\bibfnamefont {R.~F.}\
  \bibnamefont {Standaert}}, \bibinfo {author} {\bibfnamefont {J.~C.}\
  \bibnamefont {Smith}}, \bibinfo {author} {\bibfnamefont {D.~A.~A.}\
  \bibnamefont {Myles}}, \bibinfo {author} {\bibfnamefont {M.}~\bibnamefont
  {Ohl}}, \ and\ \bibinfo {author} {\bibfnamefont {J.}~\bibnamefont
  {Katsaras}},\ }\bibfield  {title} {\enquote {\bibinfo {title} {Mechanical
  properties of nanoscopic lipid domains},}\ }\href {\doibase
  10.1021/jacs.5b08894} {\bibfield  {journal} {\bibinfo  {journal} {J. Am.
  Chem. Soc.}\ }\textbf {\bibinfo {volume} {137}},\ \bibinfo {pages}
  {15772--15780} (\bibinfo {year} {2015})}\BibitemShut {NoStop}%
\bibitem [{\citenamefont {Shimobayashi}\ \emph {et~al.}(2016)\citenamefont
  {Shimobayashi}, \citenamefont {Ichikawa},\ and\ \citenamefont
  {Taniguchi}}]{Shimobayashi16}%
  \BibitemOpen
  \bibfield  {author} {\bibinfo {author} {\bibfnamefont {S.~F.}\ \bibnamefont
  {Shimobayashi}}, \bibinfo {author} {\bibfnamefont {M.}~\bibnamefont
  {Ichikawa}}, \ and\ \bibinfo {author} {\bibfnamefont {T.}~\bibnamefont
  {Taniguchi}},\ }\bibfield  {title} {\enquote {\bibinfo {title} {Direct
  observations of transition dynamics from macro- to micro-phase separation in
  asymmetric lipid bilayers induced by externally added glycolipids},}\ }\href
  {\doibase 10.1209/0295-5075/113/56005} {\bibfield  {journal} {\bibinfo
  {journal} {Europhys. Lett.}\ }\textbf {\bibinfo {volume} {113}},\ \bibinfo
  {pages} {56005} (\bibinfo {year} {2016})}\BibitemShut {NoStop}%
\bibitem [{\citenamefont {Kawakatsu}\ \emph {et~al.}(1993)\citenamefont
  {Kawakatsu}, \citenamefont {Andelman}, \citenamefont {Kawasaki},\ and\
  \citenamefont {Taniguchi}}]{Kawakatsu93}%
  \BibitemOpen
  \bibfield  {author} {\bibinfo {author} {\bibfnamefont {T.}~\bibnamefont
  {Kawakatsu}}, \bibinfo {author} {\bibfnamefont {D.}~\bibnamefont {Andelman}},
  \bibinfo {author} {\bibfnamefont {K.}~\bibnamefont {Kawasaki}}, \ and\
  \bibinfo {author} {\bibfnamefont {T.}~\bibnamefont {Taniguchi}},\ }\bibfield
  {title} {\enquote {\bibinfo {title} {Phase transitions and shapes of two
  component membranes and vesicles {I}: strong segregation limit},}\ }\href
  {\doibase 10.1051/jp2:1993177} {\bibfield  {journal} {\bibinfo  {journal} {J.
  Phys. II (France)}\ }\textbf {\bibinfo {volume} {3}},\ \bibinfo {pages}
  {971--997} (\bibinfo {year} {1993})}\BibitemShut {NoStop}%
\bibitem [{\citenamefont {Taniguchi}\ \emph {et~al.}(1994)\citenamefont
  {Taniguchi}, \citenamefont {Kawasaki}, \citenamefont {Andelman},\ and\
  \citenamefont {Kawakatsu}}]{Taniguchi94}%
  \BibitemOpen
  \bibfield  {author} {\bibinfo {author} {\bibfnamefont {T.}~\bibnamefont
  {Taniguchi}}, \bibinfo {author} {\bibfnamefont {K.}~\bibnamefont {Kawasaki}},
  \bibinfo {author} {\bibfnamefont {D.}~\bibnamefont {Andelman}}, \ and\
  \bibinfo {author} {\bibfnamefont {T.}~\bibnamefont {Kawakatsu}},\ }\bibfield
  {title} {\enquote {\bibinfo {title} {Phase transitions and shapes of two
  component membranes and vesicles {II}: weak segregation limit},}\ }\href
  {\doibase 10.1051/jp2:1994203} {\bibfield  {journal} {\bibinfo  {journal} {J.
  Phys. II (France)}\ }\textbf {\bibinfo {volume} {4}},\ \bibinfo {pages}
  {1333--1362} (\bibinfo {year} {1994})}\BibitemShut {NoStop}%
\bibitem [{\citenamefont {Taniguchi}(1996)}]{Taniguchi96}%
  \BibitemOpen
  \bibfield  {author} {\bibinfo {author} {\bibfnamefont {T.}~\bibnamefont
  {Taniguchi}},\ }\bibfield  {title} {\enquote {\bibinfo {title} {Shape
  deformation and phase separation dynamics of two-component vesicles},}\
  }\href {\doibase 10.1103/PhysRevLett.76.4444} {\bibfield  {journal} {\bibinfo
   {journal} {Phys. Rev. Lett.}\ }\textbf {\bibinfo {volume} {76}},\ \bibinfo
  {pages} {4444--4447} (\bibinfo {year} {1996})}\BibitemShut {NoStop}%
\bibitem [{\citenamefont {Amazon}\ and\ \citenamefont
  {Feigenson}(2014)}]{Amazon14}%
  \BibitemOpen
  \bibfield  {author} {\bibinfo {author} {\bibfnamefont {J.~J.}\ \bibnamefont
  {Amazon}}\ and\ \bibinfo {author} {\bibfnamefont {G.~W.}\ \bibnamefont
  {Feigenson}},\ }\bibfield  {title} {\enquote {\bibinfo {title} {Lattice
  simulations of phase morphology on lipid bilayers: Renormalization, membrane
  shape, and electrostatic dipole interactions},}\ }\href {\doibase
  10.1103/PhysRevE.89.022702} {\bibfield  {journal} {\bibinfo  {journal} {Phys.
  Rev. E}\ }\textbf {\bibinfo {volume} {89}},\ \bibinfo {pages} {022702}
  (\bibinfo {year} {2014})}\BibitemShut {NoStop}%
\bibitem [{\citenamefont {{Barrag{\'a}n Vidal}}\ \emph
  {et~al.}(2014)\citenamefont {{Barrag{\'a}n Vidal}}, \citenamefont {Rosetti},
  \citenamefont {Pastorino},\ and\ \citenamefont {M{\"u}ller}}]{Vidal14}%
  \BibitemOpen
  \bibfield  {author} {\bibinfo {author} {\bibfnamefont {I.~A.}\ \bibnamefont
  {{Barrag{\'a}n Vidal}}}, \bibinfo {author} {\bibfnamefont {C.~M.}\
  \bibnamefont {Rosetti}}, \bibinfo {author} {\bibfnamefont {C.}~\bibnamefont
  {Pastorino}}, \ and\ \bibinfo {author} {\bibfnamefont {M.}~\bibnamefont
  {M{\"u}ller}},\ }\bibfield  {title} {\enquote {\bibinfo {title} {Measuring
  the composition-curvature coupling in binary lipid membranes by computer
  simulations},}\ }\href {\doibase 10.1063/1.4901203} {\bibfield  {journal}
  {\bibinfo  {journal} {J. Chem. Phys.}\ }\textbf {\bibinfo {volume} {141}},\
  \bibinfo {pages} {194902} (\bibinfo {year} {2014})}\BibitemShut {NoStop}%
\bibitem [{\citenamefont {Olver}\ \emph {et~al.}()\citenamefont {Olver},
  \citenamefont {{Olde Daalhuis}}, \citenamefont {Lozier}, \citenamefont
  {Schneider}, \citenamefont {Boisvert}, \citenamefont {Clark}, \citenamefont
  {Miller},\ and\ \citenamefont {B.~V.~Saunders}}]{DLMF}%
  \BibitemOpen
  \bibfield  {author} {\bibinfo {author} {\bibfnamefont {F.~W.~J.}\
  \bibnamefont {Olver}}, \bibinfo {author} {\bibfnamefont {A.~B.}\ \bibnamefont
  {{Olde Daalhuis}}}, \bibinfo {author} {\bibfnamefont {D.~W.}\ \bibnamefont
  {Lozier}}, \bibinfo {author} {\bibfnamefont {B.~I.}\ \bibnamefont
  {Schneider}}, \bibinfo {author} {\bibfnamefont {R.~F.}\ \bibnamefont
  {Boisvert}}, \bibinfo {author} {\bibfnamefont {C.~W.}\ \bibnamefont {Clark}},
  \bibinfo {author} {\bibfnamefont {B.~R.}\ \bibnamefont {Miller}}, \ and\
  \bibinfo {author} {\bibfnamefont {eds.}\ \bibnamefont {B.~V.~Saunders}},\
  }\href@noop {} {\enquote {\bibinfo {title} {{NIST} {D}igital library of
  mathematical functions},}\ }\bibinfo {howpublished}
  {http://dlmf.nist.gov}\BibitemShut {NoStop}%
\end{thebibliography}%

\end{document}